\documentclass[12pt]{article}
\usepackage{jheppub}
\usepackage{amsmath}
\usepackage{amssymb}
\usepackage{amscd}
\usepackage{placeins}
\usepackage{tikz-cd}
\usepackage[export]{adjustbox}
\usepackage{subcaption}
\usetikzlibrary{snakes}
\usepackage{longtable}

\usetikzlibrary{tikzmark}
\usetikzlibrary{decorations.markings}

\normalsize

\usepackage{pslatex}
\usepackage{graphicx}
\usepackage[latin1,utf8]{inputenc}
\usepackage[T1]{fontenc}

\RequirePackage{mathtools}
\usepackage{mathdots}

\numberwithin{equation}{section}

\allowdisplaybreaks

\usepackage{xcolor}
\definecolor{cnblue}{RGB}{7,82,154}

\begin{document}

\thispagestyle{empty}

\begin{flushright}
    TUM-HEP-1552/25 \\
    ZU-TH 02/25
\end{flushright}

\vspace{1.5cm}

\begin{center}
{\Large\bf Two-loop master integrals for mixed QCD-EW \\ corrections to $gg \to H$ through $\mathcal{O}(\epsilon^2)$}\\
  \vspace{1cm}
  {\large Robin~Marzucca${}^{a}$, 
          Andrew~J.~McLeod${}^{b}$, and 
          Christoph~Nega${}^{c}$ \\
  \vspace{1cm}
      {\small \em ${}^{a}$ Physik-Institut,} 
      {\small \em Universit\"at Z\"urich,} \\[-2pt]
      {\small \em Winterthurerstrasse 190,}
      {\small \em 8057 Z\"urich, Switzerland}\\
  \vspace{2mm}
      {\small \em ${}^{b}$ Higgs Centre for Theoretical Physics,} 
      {\small \em School of Physics and Astronomy,} \\[-2pt]
      {\small \em The University of Edinburgh} 
      {\small \em Edinburgh EH9 3FD, Scotland, UK} \\
  \vspace{2mm}
      {\small \em ${}^{c}$ Physics Department,} 
      {\small \em Technical University of Munich,} \\[-2pt]
      {\small \em D - 85748 Garching, Germany}
  } 
\end{center}

\vspace{2cm}

\begin{abstract}\noindent
We consider mixed strong-electroweak corrections to Higgs production via gluon fusion, in which the Higgs boson couples to the top quark. Using the method of differential equations, we compute all of the master integrals that contribute to this process at two loops through $\mathcal{O}(\epsilon^2)$ in the dimensional regularization parameter $\epsilon = (d-4)/2$, keeping full analytic dependence on the top quark, Higgs, W, and Z boson masses. We present the results for these master integrals in terms of iterated integrals whose kernels depend on elliptic curves.
\end{abstract}
\vspace*{\fill}

\newpage



\section{Introduction}
\label{sec:intro}

Higgs boson production via gluon fusion is one of the central observables at the LHC. The importance of this channel derives not only from it being the dominant mode of Higgs production at LHC energies but also from its sensitivity to particles that simultaneously couple to gluons and receive large masses from the Higgs mechanism. Due to this sensitivity, gluon fusion measurements provide valuable input for improving our measurements of Standard Model parameters, while also allowing us to probe for signals of beyond-the-Standard-Model physics.

Given its importance, it is no surprise that a great deal of effort has gone into improving the theoretical precision of Standard Model predictions for $gg \to H$ over the last two decades (see for instance~\cite{Aglietti:2004nj, Degrassi:2004mx, Actis:2008ug, Anastasiou:2008tj, Baikov:2009bg, Gehrmann:2010ue, Anastasiou:2014vaa, Anastasiou:2015vya, Anastasiou:2016cez, Bonetti:2017ovy, Anastasiou:2018adr, Becchetti:2020wof, Czakon:2021yub}). However, despite the impressive progress that has been made, further reductions to the theoretical uncertainty of Standard Model predictions must be made to match the projected experimental precision to be achieved by the high-luminosity LHC~\cite{Cepeda:2019klc}. One of the sources of this theoretical uncertainty comes from not having full control over the finite-mass effects associated with top quarks and electroweak bosons in mixed quantum chromodynamic-electroweak (QCD-EW) corrections. In particular, while the two-loop mixed QCD-EW effects involving light virtual quarks were computed over twenty years ago~\cite{Aglietti:2004nj}, contributions coming from virtual top quarks have still only been evaluated numerically~\cite{Actis:2008ug} or as expansions around kinematic limits~\cite{Degrassi:2004mx}.\footnote{Similarly, three-loop mixed QCD-EW corrections have only been computed for massless quarks~\cite{Anastasiou:2008tj, Bonetti:2017ovy, Anastasiou:2018adr}.} An analytic calculation of $gg \to H$ that incorporates the full dependence of the Higgs mass, top quark mass, and electroweak boson masses at two loops is thus long overdue.

In this paper, we make progress towards this goal by computing all of the Feynman integrals that contribute to $gg \to H$ at $\mathcal{O}(g^3 g_s^2)$ in which the Higgs boson couples to the top quark, where $g_s$ and $g$ are the strong and electroweak coupling constants. These contributions appear at two loops, and we compute them in dimensional regularization $(d = 4 - 2 \epsilon)$ through $\mathcal{O}(\epsilon^2)$, keeping all dependence on the top quark, Higgs, and electroweak boson masses. In this way, we pave the way for analytically determining $gg \to H$ not only at next-to-leading-order (NLO) but also next-to-next-to-leading-order (NNLO), with full dependence on the heaviest Standard Model particles. Along with improving Standard Model predictions across a broad range of energy scales, achieving this level of precision will provide valuable insight into the mathematical features of Standard Model amplitudes and the extent to which these amplitudes exhibit the same types of analytic, geometric, and number-theoretic structures that have been observed in massless theories~\cite{Panzer:2016snt, Schnetz:2017bko, Caron-Huot:2018dsv, Caron-Huot:2019bsq, Abreu:2021smk, Dixon:2021tdw, Dixon:2022xqh}.

One of the difficulties that quickly arises when computing amplitudes that involve heavy virtual particles is the appearance of special functions that go beyond iterated integrals~\cite{Chen} involving only $d\log$ forms---a space of functions which is nearly coextensive with multiple polylogarithms~\cite{G91b, Goncharov:1998kja, Remiddi:1999ew, Borwein:1999js, Moch:2001zr}.\footnote{While iterated integrals of $d\log$ forms do not always evaluate to multiple polylogarithms~\cite{Duhr:2020gdd}, all the examples of this type we encounter in this paper can be expressed in terms of these functions.} Even in two- and three-particle kinematics, integrals over more complicated integration kernels---which involve elliptic curves or even Calabi-Yau manifolds---are known to start appearing at two loops~\cite{SABRY1962401, Huang:2013kh,Vanhove:2014wqa, Bourjaily:2017bsb, Bourjaily:2018yfy, Honemann:2018mrb, Bourjaily:2019hmc,Jockers:2020sdr,Bourjaily:2022bwx, Duhr:2022dxb, Duhr:2022pch, Marzucca:2023gto, Forner:2024ojj, Duhr:2024bzt, Duhr:2024hjf}.\footnote{Calabi-Yau manifolds have also been observed in the treatment of the two-body scattering problem in classical general relativity~\cite{Bern:2021yeh, Frellesvig:2023bbf, Klemm:2024wtd, Driesse:2024feo}.} The complication from elliptic curves indeed also arises for $gg \to H$ when virtual top quarks and electroweak bosons are included due to the appearance of the sunrise integral with all nonzero internal masses~\cite{Laporta:2004rb, Adams:2013nia, Bloch:2013tra, Remiddi:2013joa, Adams:2014vja, Adams:2015gva, Adams:2015ydq, Bloch:2016izu, Broedel:2017siw, Bogner:2017vim, Adams:2018yfj}. While this fact has long stalled the analytic evaluation of these amplitudes, considerable progress has recently been made in understanding how to work with these classes of elliptic functions (see for instance~\cite{Adams:2015ydq, Adams:2017ejb, Broedel:2018iwv, Broedel:2018qkq, Duhr:2019rrs, Walden:2020odh}). In particular, systematic methods for putting the differential equations that describe elliptic families of Feynman integrals~\cite{Kotikov:1990kg, Kotikov:1991pm, Remiddi:1997ny, Gehrmann:1999as} into canonical form~\cite{Henn:2013pwa} have been studied by a number of authors~\cite{Adams:2017tga, Adams:2018yfj, Dlapa:2020cwj, Dlapa:2022wdu, Jiang:2023jmk, Gorges:2023zgv}. We will make use of these strategies here.

We begin our calculation by identifying the complete set of two-loop mixed QCD-EW Feynman integrals that contribute to $gg \to H$, in which the outgoing Higgs couples to a virtual top quark. We then reduce these integrals to five families of master integrals and---using the strategies referred to above---put the systems of differential equations that govern these families into canonical form. To determine the boundary conditions, we require that no pseudo-threshold singularities arise in physical kinematics; any boundary conditions that are not fixed by this condition are then determined using the method of regions~\cite{Smirnov:1990, Smirnov:1994tg, BENEKE1998321, Smirnov:1998vk, Pak:2010pt, Smirnov:2012gma, Heinrich:2021dbf, Ma:2023hrt}. In this way, we arrive at iterated integral representations for each of our master integrals.

In fact, three of the families of integrals we encounter can be expressed in terms of iterated integrals involving $d\log$ integration kernels to all orders in $\epsilon$. We present these integrals in terms of multiple polylogarithms---using the notation found, for instance, in~\cite{Vollinga:2004sn, Duhr:2019tlz}---through weight six. The remaining two families are of elliptic type due to the appearance of the sunrise integral with two unequal masses. Since only a single elliptic curve appears in these integrals, we expect that they should be expressible in terms of elliptic multiple polylogarithms~\cite{brownLevin, Broedel:2014vla, Adams:2015ydq, Adams:2016xah, Broedel:2017kkb}. However, we here simply present these integrals as iterated integrals whose integration kernels involve elliptic periods related to the sunrise elliptic curve, leaving the task of expressing these integrals in terms of elliptic multiple polylogarithms to future work. Our iterated integral representation still allows for the use of numerical evaluation methods for iterated integrals, such as those that have been implemented in {\sc GiNaC}~\cite{Bauer:2000cp, Vollinga:2004sn, Walden:2020odh}, or that make use of generalized power series expansions~\cite{Moriello:2019yhu, Hidding:2020ytt}.

We have numerically cross-checked our iterated integral expressions for each master integral, comparing to direct numerical evaluation of the original integral using AMFlow~\cite{Liu:2017jxz,Liu:2022chg}. In the families that can be expressed in terms of multiple polylogarithms, this can be easily done by numerically evaluating our expressions using {\sc GiNaC}~\cite{Bauer:2000cp}. For the families that involve elliptic integration kernels, we carried out this cross-check by first series expanding the elliptic kernels with respect to $s$, such that the iterated integrations can be performed analytically; we then evaluated these expansions numerically.

This paper is structured as follows: In section~\ref{sec:gg_H}, we describe the physical process we consider in more detail, along with the procedure we used to identify the complete set of two-loop integrals that arise in its calculation, and how we reduced these integrals to a basis. In section~\ref{sec:masters}, we outline our choice of master integrals and comment on their analytic calculation in terms of iterated integrals. In section~\ref{sec:conclusion}, we conclude the paper and outline the next logical steps to take. 

We have also included an appendix and ancillary files. Appendix~\ref{sec:basis_description} details our choice of master integrals, before we have rotated to a canonical basis. These definitions are also provided in ancillary files, alongside the analytic expressions for the canonical master integrals and the rotation used to get from our initial basis to the canonical one. We have, in addition, provided a sample {\sc Mathematica} notebook that illustrates how to navigate the data that we have included in the ancillary files.

\section{Conventions and Definitions}
\label{sec:gg_H}

We are interested in computing the complete set of two-loop Feynman integrals that appear in mixed QCD-EW contributions to $gg \to H$ when the outgoing Higgs couples to a virtual top quark. In order to identify these integrals, we first generate all such two-loop graphs that can appear at $\mathcal{O}(g^3 g_s^2)$ using {\sc QGRAF}~\cite{Nogueira:1991ex}, and dress these graphs with the appropriate Feynman rules using {\sc FORM}~\cite{Vermaseren:2000nd, Kuipers:2012rf, Ruijl:2017dtg} and {\sc Mathematica}~\cite{Mathematica}. Our conventions for the Standard Model Feynman rules match the Universal FeynRules Output~\cite{Degrande:2011ua}, as found, for instance, in the {\sc tapir} code~\cite{Gerlach:2022qnc}. We specifically retain the gauge dependence of the vector boson propagators, ensuring that the cancellation of the gauge parameters can serve as a cross-check when the amplitude is computed.

For an off-shell Higgs boson, the diagrams that contribute to $gg \to H$ depend on the Mandelstam variable
\begin{equation}
s = (p_1+ p_2)^2 \, ,
\end{equation} 
where $p_1$ and $p_2$ are the incoming momenta of the gluons. We also keep the full analytic dependence on the Higgs mass $m_H$, the $W$ and $Z$ boson masses $m_W$ and $m_Z$, and the top quark mass $m_t$. All other quark masses are set to zero. Therefore, we expect the transcendental functions that appear in our calculation to depend on the four dimensionless ratios 
\begin{equation} \label{eq:vars}
	\smash{\bigg\{\frac{s}{m^2_H}, \frac{s}{m^2_t}, \frac{s}{m^2_W}, \frac{s}{m^2_Z}\bigg\}} \, .
\end{equation}
In the EW corrections that we consider to the leading-order QCD contribution, however, no individual diagram can depend on more than two of these masses. As a result, the transcendental part of each of our integrals depends on at most two of the variables in~\eqref{eq:vars}. 

Due to gauge invariance, only a single form factor contributes to $gg \to H$. As a result, the amplitude can be written as
\begin{equation}
\mathcal{M}^{c_1 c_2}_{\lambda_1 \lambda_2} = \mathcal{F}(s, m_H, m_t, m_W, m_Z) \, \delta^{c_1 c_2}\, \epsilon_{\lambda_1}(p_1) \, \epsilon_{\lambda_2}(p_2) \, ,
\end{equation}
where $c_i$ and $\lambda_i$ represent the color index and helicity associated with gluon $i$, and we have the constraint that $\epsilon_{\lambda_i}(p_i) \cdot p_i = 0$. The contribution of each Feynman integral to this form factor can be extracted using the projection operator~\cite{Bonetti:2017ovy} 
\begin{equation}
\mathcal{P}^{\mu \nu}_{c_1 c_2} = \frac{\delta_{c_1 c_2}}{\left(N_c^2-1\right) \left(d-2\right)} \left(-g^{\mu \nu} + \frac{p_1^\mu p_2^\nu + p_1^\nu p_2^\mu}{p_1 \cdot p_2} \right) \, ,
\end{equation}
where $N_c$ is the number of colors, $d = 4 - 2 \epsilon$ is the spacetime dimension, and the $c_i$ are again color indices (in the adjoint representation). We apply this projection operator to our expression for the amplitude, carrying out all the relevant color and Dirac algebra using {\sc FORM}. Because we are just considering leading-order EW corrections to the QCD amplitude, at most one electroweak gauge boson appears in each Feynman diagram (and when these gauge bosons do appear, both ends of the corresponding propagator attach to the same fermion loop). As a result, we can distinguish between three possible outcomes after expanding the expressions. 
\begin{enumerate}
\item A term contains no $\gamma_5$. This case is trivial.
\item A term contains two $\gamma_5$'s. Since both occurrences of $\gamma_5$ originate from a coupling to the same fermion loop, we can eliminate $\gamma_5$ by anti-commuting until we arrive at $\gamma_5^2 = 1$.
\item A term contains only one $\gamma_5$. In this case, we can not eliminate $\gamma_5$ in a trivial way, and the trace in Dirac space will generate a Levi-Civita tensor.
\end{enumerate}
We see that, for the diagrams considered in this work, each term will feature at most one Levi-Civita tensor. Since the process only depends on two independent momenta, this means that all contributions involving the axial couplings will be annihilated by the projection operator~$\mathcal{P}^{\mu \nu}_{c_1 c_2} $.

\begin{figure}[t]
\centering
\begin{subfigure}{0.32\textwidth}
\scalebox{.82}{
\begin{tikzpicture}[baseline=(current bounding box.center), line width=2.2, scale=1,line cap=round]
    \draw[black] (-2,-1) -- node[left] {$m_t$} (-2,1) -- node[above] {$m_t$} (0,1);
    \draw[black] (2.1,0) -- node[below,xshift=0.3cm,yshift=.05cm] {$m_t$} (0,-1) -- node[below] {$m_t$} (-2,-1) -- (-2,1);
    \draw[black] (0,1) to node[above,xshift=0.3cm,yshift=-.1cm]{$m_t$} (2.1,0);
    \draw[black, line width=1] (0,1) -- node[left] {} (0,-1);
    \draw[black, line width=1] (-2,-1) -- (-2.85,-1.85);
    \draw[black, line width=1] (-2,1) -- (-2.85,1.85);
    \draw[black,dashed] (2.1,0) -- (3.3,0);
	\node [black] at (-2.3,-1.8) {$p_1$};
	\node [black] at (-2.3,1.8) {$p_2$};
	\node [black] at (3,-.3) {$p_3$};
\end{tikzpicture}} \vspace{-.16cm}
\caption{Family 1}
\end{subfigure} 
\hfill
\begin{subfigure}{0.6\textwidth}
\tikzmark{fam2L} \scalebox{.82}{
\begin{tikzpicture}[baseline=(current bounding box.center), line width=2.2, scale=1,line cap=round]
    \draw[black] (-2,-1) -- node[left] {$m_t$} (-2,1) -- node[above] {$m_t$} (0,1);
    \draw[black] (2.1,0) -- node[below,xshift=0.3cm,yshift=.05cm] {$m_t$} (0,-1) -- node[below] {$m_t$} (-2,-1) -- (-2,1);
    \draw[black] (0,1) to node[above,xshift=0.3cm,yshift=-.1cm]{$m_t$} (2.1,0);
    \draw[black] (0,1) -- node[left] {$M$} (0,-1);
    \draw[black, line width=1] (-2,-1) -- (-2.85,-1.85);
    \draw[black, line width=1] (-2,1) -- (-2.85,1.85);
    \draw[black,dashed] (2.1,0) -- (3.3,0);
	\node [black] at (-2.3,-1.8) {$p_1$};
	\node [black] at (-2.3,1.8) {$p_2$};
	\node [black] at (3,-.3) {$p_3$};
\end{tikzpicture} \qquad 
\begin{tikzpicture}[baseline=(current bounding box.center), line width=2.2, scale=1,line cap=round]
    \draw[black] (-2,-1) -- node[left] {$m_t$} (-2,1) -- node[above] {$m_t$} (0,1);
    \draw[black] (0,-1) -- node[below] {$m_t$} (-2,-1) -- (-2,1);
    \draw[black, line width=1] (-2,-1) -- (-2.85,-1.85);
    \draw[black, dashed] (-2,1) -- (-2.85,1.85);
    \draw[black, line width=1] (0,1) -- (.85,1.85);
    \draw[black, line width=1] (0,-1) to [bend left] node[right,xshift=.1cm,yshift=-.1cm] {} (0,1);
    \draw[black] (0,-1) to [bend right] node[right,xshift=.1cm,yshift=0cm] {$M$} (0,1);
	\node [black] at (-2.3,-1.8) {$p_1$};
	\node [black] at (-2.3,1.8) {$p_3$};
	\node [black] at (.3,1.8) {$p_2$};
\end{tikzpicture}}\tikzmark{fam2R} \vspace{-.16cm}
\caption{Family 2}
\end{subfigure} 
\\[.4cm]
\begin{subfigure}{0.32\textwidth}
\scalebox{.82}{
\begin{tikzpicture}[baseline=(current bounding box.center), line width=2.2, scale=1,line cap=round]
    \draw[black,line width=1] (-2,-1) -- node[left] {} (-2,1) -- node[above] {} (0,1);
    \draw[black] (2.1,0) -- node[below,xshift=0.3cm,yshift=.05cm] {$m_t$} (0,-1);
    \draw[black,line width=1] (0,-1) -- node[below] {} (-2,-1) -- (-2,1);
    \draw[black] (0,1) to node[above,xshift=0.3cm,yshift=-.1cm]{$m_t$} (2.1,0);
    \draw[black] (0,1) -- node[left] {$M$} (0,-1);
    \draw[black, line width=1] (-2,-1) -- (-2.85,-1.85);
    \draw[black, line width=1] (-2,1) -- (-2.85,1.85);
    \draw[black,dashed] (2.1,0) -- (3.3,0);
	\node [black] at (-2.3,-1.8) {$p_1$};
	\node [black] at (-2.3,1.8) {$p_2$};
	\node [black] at (3,-.3) {$p_3$};
\end{tikzpicture}} \vspace{-.16cm}
\caption{Family 3}
\end{subfigure}
\qquad \qquad
\begin{subfigure}{0.32\textwidth}
\scalebox{.82}{
\begin{tikzpicture}[baseline=(current bounding box.center), line width=2.2, scale=1,line cap=round]
    \draw[black] (-2,-1) -- node[left] {$m_t$} (-2,1) -- node[above] {$m_t$} (0,1);
    \draw[black] (2.1,0) -- node[below,xshift=0.3cm,yshift=.05cm] {$m_t$} (0,-1) -- node[below] {$m_t$} (-2,-1) -- (-2,1);
    \draw[black] (0,1) to node[above,xshift=0.3cm,yshift=-.1cm]{$m_t$} (2.1,0);
    \draw[black, line width=1] (0,1) -- node[left] {} (0,-1);
    \draw[black, line width=1] (-2,-1) -- (-2.85,-1.85);
    \draw[black, dashed] (-2,1) -- (-2.85,1.85);
    \draw[black, line width=1] (2.1,0) -- (3.3,0);
	\node [black] at (-2.3,-1.8) {$p_1$};
	\node [black] at (-2.3,1.8) {$p_3$};
	\node [black] at (3,-.3) {$p_2$};
\end{tikzpicture}} \vspace{-.16cm}
\caption{Family 4}
\end{subfigure}
\\[.4cm] 
\begin{subfigure}{.99\textwidth} \centering
\tikzmark{fam5L} \scalebox{.82}{
\begin{tikzpicture}[baseline=(current bounding box.center), line width=2.2, scale=1,line cap=round]
    \draw[black] (-2,-1) -- node[left] {$m_t$} (-2,1) -- node[above] {$m_t$} (0,1);
    \draw[black] (0,-1) -- node[below] {$m_t$} (-2,-1) -- (-2,1);
    \draw[black, line width=1] (-2,-1) -- (-2.85,-1.85);
    \draw[black, dashed] (-2,1) -- (-2.85,1.85);
    \draw[black, line width=1] (0,1) -- (.85,1.85);
    \draw[black] (0,-1) -- node[right] {$m_t$} (0,1);
    \draw[black] (-2,1) -- node[right] {$M$} (0,-1);
	\node [black] at (-2.3,-1.8) {$p_1$};
	\node [black] at (-2.3,1.8) {$p_3$};
	\node [black] at (.3,1.8) {$p_2$};
\end{tikzpicture} \qquad 
\begin{tikzpicture}[baseline=(current bounding box.center), line width=2.2, scale=1,line cap=round]
    \draw[black] (-2,-1) -- node[left] {$m_t$} (-2,1) -- node[above] {$m_t$} (0,1);
    \draw[black] (0,-1) -- node[below] {$m_t$} (-2,-1) -- (-2,1);
    \draw[black, line width=1] (-2,-1) -- (-2.85,-1.85);
    \draw[black, line width=1] (-2,1) -- (-2.85,1.85);
    \draw[black, dashed] (0,1) -- (.85,1.85);
    \draw[black, line width=1] (0,-1) to [bend left] node[right,xshift=.1cm,yshift=-.1cm] {} (0,1);
    \draw[black] (0,-1) to [bend right] node[right,xshift=.1cm,yshift=0cm] {$M$} (0,1);
	\node [black] at (-2.3,-1.8) {$p_1$};
	\node [black] at (-2.3,1.8) {$p_2$};
	\node [black] at (.3,1.8) {$p_3$};
\end{tikzpicture} \quad 
\begin{tikzpicture}[baseline=(current bounding box.center), line width=2.2, scale=1,line cap=round]
    \draw[black] (-2,-1) -- node[left] {$m_t$} (-2,1) -- node[above] {$m_t$} (0,1);
    \draw[black] (0,-1) -- node[below] {$m_t$} (-2,-1) -- (-2,1);
    \draw[black, line width=1] (-2,-1) -- (-2.85,-1.85);
    \draw[black, dashed] (-2,1) -- (-2.85,1.85);
    \draw[black, line width=1] (0,1) -- (.85,1.85);
    \draw[black] (0,-1) to [bend left] node[right,xshift=-.74cm,yshift=0cm] {$m_t$} (0,1);
    \draw[black] (0,-1) to [bend right] node[right,xshift=.1cm,yshift=0cm] {$M$} (0,1);
	\node [black] at (-2.3,-1.8) {$p_1$};
	\node [black] at (-2.3,1.8) {$p_3$};
	\node [black] at (.3,1.8) {$p_2$};
\end{tikzpicture}} \tikzmark{fam5R} \vspace{-.16cm}
\begin{tikzpicture}[overlay, remember picture,decoration={markings,mark=at position .99 with {\arrow[scale=1.4,>=stealth]{>}}}]
    \draw [dashed,rounded corners=.4cm] ([yshift=1.76cm,xshift=-.1cm]{pic cs:fam2L}) rectangle ([yshift=-1.7cm,xshift=.1cm]{pic cs:fam2R});
    \draw [dashed,rounded corners=.4cm] ([yshift=1.76cm,xshift=-.1cm]{pic cs:fam5L}) rectangle ([yshift=-1.7cm,xshift=.1cm]{pic cs:fam5R});
\end{tikzpicture}
\caption{Family 5}
\end{subfigure}
\caption{The five two-loop integral families that appear at $\mathcal{O}(g^3 g_s^2)$ in $gg \to H$, in which the Higgs boson couples to a top quark. The mass $M$ can take the values $m_H$, $m_W$, or $m_Z$. Thin lines represent massless particles, and the dashed line represents the outgoing Higgs boson. 
Note that families 2 and 5 involve multiple top-level topologies.}
\label{fig:master_topology}
\end{figure}
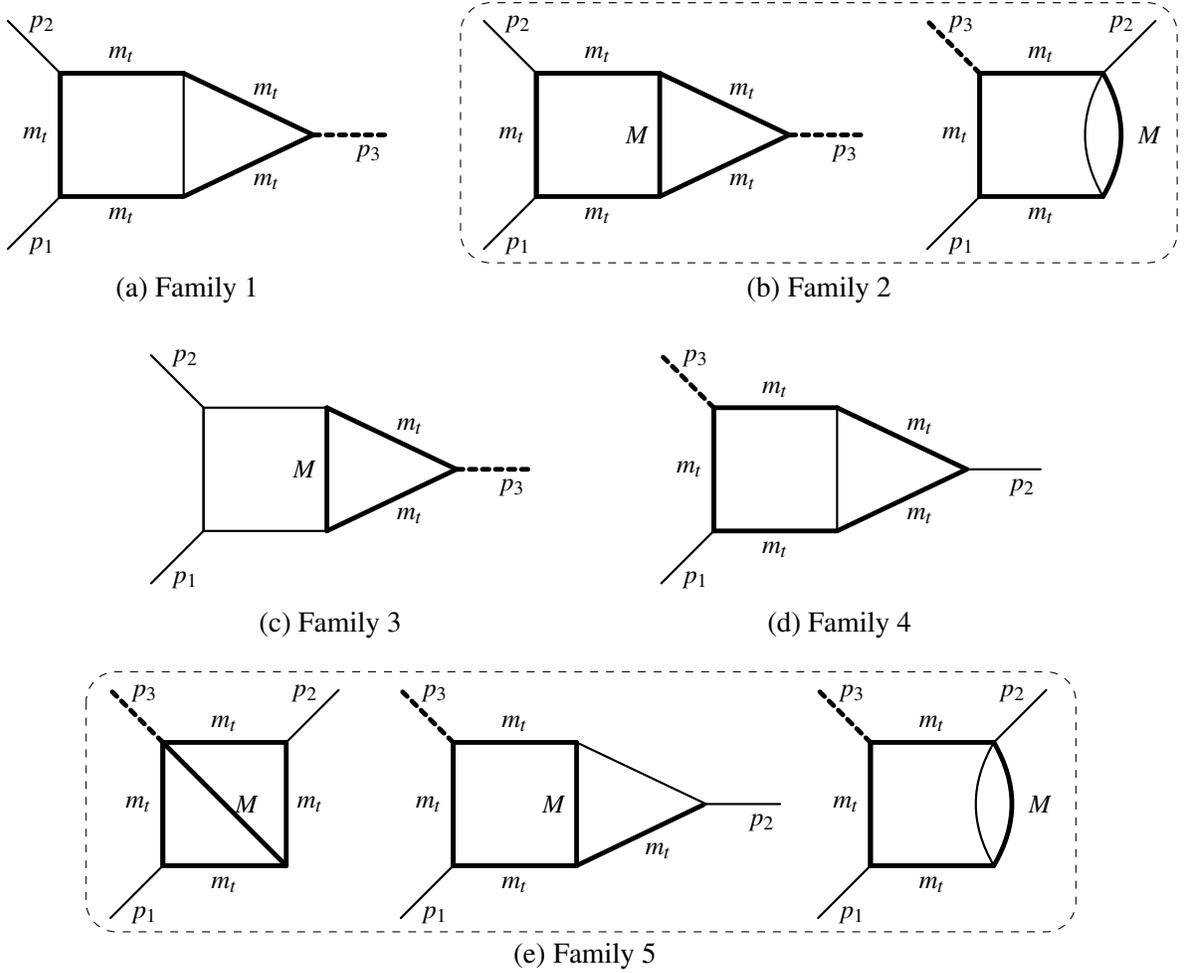

Finally, we collect all the scalar Feynman integrals that remain in our expression and feed them into {\sc Reduze}~\cite{vonManteuffel:2012np}. Loop momentum shifts are then used to group together integrals that draw upon a shared set of propagators, and kinematic numerators are similarly expressed in terms of inverse propagators. This allows us to reduce these integrals to five families of master integrals using {\sc Kira}~\cite{Klappert:2020nbg}. The top-level topology for each of these families of integrals is depicted in Figure~\ref{fig:master_topology}. Note that multiple top sectors are grouped together in families 2 and 5.\footnote{We thank the referee for pointing out that, by choosing more carefully the ISPs, one could have collected all top sectors into even fewer families.} 
The basis of master integrals we consider in each of these families is specified in Appendix~\ref{sec:basis_description}. While $m_t$ appears in every integral (since we are considering the contribution to $gg \to H$ in which the Higgs couples to a virtual top quark), the second mass $M$ that appears in families 2, 3, and 5 can take any of the values $m_H$, $m_W$, or $m_Z$ (or the gauge-dependent values of the corresponding Goldstone bosons).

\section{Master Integrals}
\label{sec:masters}

In $d = 4-2\epsilon$ dimensions, the scalar integrals identified in the last section take the form
\begin{equation}
I^{(f)}_{a_1,\dots,a_E} = e^{2\epsilon\gamma} \int   \frac{\mathrm d^d\ell_1}{i \pi^{d/2}}\frac{\mathrm d^d\ell_2}{i \pi^{d/2}}
 \prod_{i=1}^E \left(D_i^{(f)}\right)^{-a_i} \, ,
\end{equation}
where $\smash{D_i^{(f)}}$ represents the $i^{\text{th}}$ propagator in family $f$, which is raised to the (possibly negative) power $a_i$, and $\gamma$ is the Euler-Mascheroni constant.\footnote{Note that our conventions match those of AMFlow, modulo the factor $e^{2\epsilon\gamma}$~\cite{Liu:2022chg}.} Our five families are defined by the propagators 
\begin{align}
D^{(1)}_i &= \{\, k_1^2 - m_t^2\, ,\, (k_1 + p_1)^2 - m_t^2\, ,\, (k_1 - p_2)^2 - m_t^2\, ,\, \label{eq:props_1} \\ 
&\qquad(k_1 + k_2)^2\, ,\, (k_2 - p_1)^2 - m_t^2\, ,\, (k_2 + p_2)^2 - m_t^2\, ,\, k_2^2 \,\} \, , \nonumber \\
D^{(2)}_i &= \{\, k_1^2 - m_t^2\, ,\, (k_1 + p_1)^2 - m_t^2\, ,\, (k_1 - p_2)^2 - m_t^2\, ,\, \\ 
&\qquad(k_1 + k_2)^2 - M^2\, ,\, (k_2 - p_1)^2 - m_t^2\, ,\, (k_2 + p_2)^2 - m_t^2\, ,\, k_2^2 \,\} \, , \nonumber \\
D^{(3)}_i &= \{\, k_1^2\, ,\, (k_1 + p_1)^2\, ,\, (k_1 - p_2)^2\, ,\, (k_1 + k_2)^2 - M^2\, ,\, \\
&\qquad(k_2 - p_1)^2 - m_t^2\, ,\, (k_2 + p_2)^2 - m_t^2\, ,\, k_2^2 \,\} \, , \nonumber \\
D^{(4)}_i &= \{\, k_1^2 - m_t^2\, ,\, (k_1 + p_1)^2 - m_t^2\, ,\, (k_1 - p_2)^2 - m_t^2\, ,\, \\ 
&\qquad(k_1 + k_2)^2\, ,\, (k_2 - p_1)^2\, ,\, (k_2 + p_2)^2 - m_t^2\, ,\, k_2^2 - m_t^2 \,\} \, , \nonumber \\
D^{(5)}_i &= \{\, k_1^2 - m_t^2\, ,\, (k_1 + p_1)^2 - m_t^2\, ,\, (k_1 - p_2)^2 - m_t^2\, ,\,  \label{eq:props_5} \\ 
&\qquad(k_1 + k_2)^2 - M^2\, ,\, (k_2 - p_1)^2\, ,\, (k_2 + p_2)^2 - m_t^2\, ,\, k_2^2 - m_t^2 \,\} \, , \nonumber
\end{align}
where the external momenta $p_1$ and $p_2$ are both incoming, and $k_1$ and $k_2$ represent loop momenta. There are 10, 30, 15, 5, and 26 master integrals in families 1 through 5, respectively.

We choose to solve the different integral families individually, as they exhibit varying degrees of complexity, allowing us to express the simpler integral families in terms of simpler classes of functions. To solve the systems of differential equations that describe each of these families of integrals, we rotate each family into a canonical basis such that the individual differential equations take the following form 
\begin{equation}
\mathrm d \vec{I} = \epsilon \sum_i \omega_i \, A_i \cdot \vec{I} \label{eq:canonical_form} \, ,
\end{equation}
where the $\omega_i$ are differential one-forms, and the $A_i$ are rational numeric matrices, both of which are independent of $\epsilon$. For the families of integrals that evaluate to multiple polylogarithms, the rotations needed to put the differential equations in this form can be found by analyzing leading singularities or using publicly available software (see for instance~\cite{Prausa:2017ltv, Gituliar:2017vzm, Meyer:2017joq, Dlapa:2020cwj, Lee:2020zfb}). Although no software packages exist for rotating elliptic families of Feynman integrals into canonical form, strategies for accomplishing this task have also been proposed~\cite{Adams:2018yfj, Dlapa:2020cwj, Dlapa:2022wdu, Jiang:2023jmk, Gorges:2023zgv}. In this manuscript, we follow the approach introduced in~\cite{Gorges:2023zgv}, which involves analyzing the elliptic period integrals that describe the maximal cuts of the corresponding elliptic master integrals. More specifically, for sectors that have elliptic maximal cuts, one considers the corresponding fundamental matrix of solutions, also known as the Wronskian matrix, which collects all maximal cuts of the sector~\cite{Primo:2016ebd, Primo:2017ipr}. The main part of the method introduced in~\cite{Gorges:2023zgv} consists of splitting this Wronskian matrix into a semi-simple part that contains the generalized algebraic parts of the matrix, and a unipotent piece that contains the transcendental or logarithmic contributions to the maximal cuts. Multiplying the pre-canonical basis with the inverse of the semi-simple part allows us to strip off the generalized leading singularities and, after readjusting the transcendental weights of the integrals and integrating out non-$\epsilon$-factorized terms, yields an $\epsilon$-factorized differential equation. After this step, to obtain an $\epsilon$-factorized differential equation beyond the maximal cut, one can easily integrate out non-$\epsilon$-factorized contributions in the sub-sectors. For sectors whose maximal cuts are algebraic, but that couple to one of the elliptic sectors, we first use standard methods to bring the polylogarithmic parts into $\epsilon$-factorized form. Then, one can again integrate out non-$\epsilon$-factorized elliptic contributions. In this step, it may be necessary to introduce new functions, namely integrals over elliptic periods, to bring the whole system into canonical form. 

The above method works particularly well if the pre-canonical basis for the elliptic sectors is chosen carefully. In the elliptic sectors encountered in this paper, we have chosen our master integrals so that they are related to the standard differential forms on an elliptic curve, namely the differential forms of the first, second, and third kinds. In the end, following this approach and using the standard techniques for polylogarithms, we have been able to bring the differential equations that describe all of the integral families into the canonical form shown in~\eqref{eq:canonical_form}.

Once we have found a canonical basis of master integrals, we can integrate the corresponding differential equation order by order in $\epsilon$. Doing so fully captures the kinematic dependence of the master integrals up to (transcendental) constants, which can be fixed by calculating the master integrals at a single phase space point. We determine these constants in the $s \rightarrow 0$ limit, relying on the fact that the integrals that do not have any massless threshold singularities are regular in this limit. Conversely, integrals such as those in family 3 that contain massless bubbles exhibit a threshold for $s \rightarrow 0$. However, their boundary values can still be computed using expansions by regions~\cite{Smirnov:1990, Smirnov:1994tg, BENEKE1998321, Smirnov:1998vk, Pak:2010pt, Smirnov:2012gma, Heinrich:2021dbf, Ma:2023hrt}, where we approach the $s \rightarrow 0$ limit from below. This allows us to derive iterated integral representations of our master integrals that are first valid in the Euclidean region, namely $s < 0$, and by analytic continuation also beyond. All boundary values can be expressed as linear combinations of one- and two-loop tadpole integrals with up to two different masses (which can be found, for instance, in~\cite{DAVYDYCHEV1993123}).

\subsection{Families 1 and 4}

The first and fourth families are the easiest to compute, as they only feature diagrams which already appeared in the pure QCD corrections to this process, and which depend only on $s$ and $m_t$; they have previously been computed, for instance in~\cite{Spira:1995rr, Harlander:2005rq, Anastasiou:2006hc, Aglietti:2006tp}. The only singularities that appear in these integrals are the $s \to 4 m_t^2$ threshold and $s \to 0$ pseudo-threshold. In particular, after the differential equations for these families are put in canonical form, it can be seen that both logarithmic and algebraic branch cuts arise at each of these kinematic points. In the connection matrix $\sum_i \omega_i \, A_i$, the algebraic branch cuts always arise in the forms $\omega_i$ together as the algebraic factor
\begin{equation}
\sqrt{-s\vphantom{\bar{l}_t}} \sqrt{\vphantom{\bar{l}_t} \smash{4 m_t^2 - s}} \, ,
\end{equation}
which can be rationalized using the change of variables 
\begin{equation}
\frac{s}{m_t^2} \to \frac{4 \, x^2}{x^2 - 1} \, .
\end{equation}
After this rationalization, both families of integrals can be seen to evaluate to multiple polylogarithms with symbol alphabet $\{x, 1+x, 1-x\}$.

As stated above, we fix the integration constants for our differential equations in the $s\to0$ limit. This is a regular limit, as can be seen by inspecting the corresponding Feynman integrals (see sections \ref{subsec:F1} and \ref{subsec:F4}) since they do not have a massless cut. Moreover, it is expected that the pseudo-threshold should not be accessible on the physical sheet. Using the method of regions~\cite{Smirnov:1990, Smirnov:1994tg, BENEKE1998321, Smirnov:1998vk, Pak:2010pt, Smirnov:2012gma, Heinrich:2021dbf, Ma:2023hrt}, it can be checked that only a single expansion region contributes in each of the master integrals (the so-called hard region). Correspondingly, the leading term in this limit can be computed by simply setting the Mandelstam variable $s$ to zero. The resulting integrals can be easily reduced to tadpole integrals and computed as functions of $\epsilon$.

\subsection{Family 3}

The third family depends on two internal masses, so the iterated integrals it gives rise to depend on two dimensionless variables. One can see from Figure~\ref{fig:master_topology} that the massive sunrise integral does not appear in this family, so we expect it can be evaluated in terms of multiple polylogarithms. However, we also note the appearance of a new product of square roots
\begin{equation}
\sqrt{s-(M - m_t)^2} \sqrt{s-(M + m_t)^2}\, ,
\end{equation}
which appears in addition to $\sqrt{-s\vphantom{\vec{{t_t}}}}\sqrt{4m_t^2-s}$. To rationalize both algebraic factors, we make the change of variables
\begin{equation}
\frac{s}{m_t^2} \to \frac{4 x^2}{x^2-1}\, , \qquad \frac{M^2}{m_t^2} \to \frac{z \left(3 x^2-4 x z+1\right)}{\left(x^2-1\right) (x-z)}  \, .
\end{equation}
Then, we can integrate the differential equation order by order in $\epsilon$. The resulting expressions for the master integrals are pure, uniformly transcendental linear combinations of polylogarithms $G_{\vec{a}}(z)$ and $G_{\vec{b}}(x)$, where
\begin{align} \label{eq:a_indices}
a_i \in &\Big\lbrace 0, x,  \frac{1}{2}(x \pm 1), \frac{1}{4}(3x+x^{-1}),x \pm \frac{1}{2} \sqrt{x^2-1}, \\ \nonumber
& \frac{1}{2} \left(x + 1 \pm \sqrt{5 x^2 + 6x + 5}\right), \frac{1}{2} \left(x - 1 \pm \sqrt{5 x^2 + 6x + 5}\right)  \Big\rbrace \, , \\
b_i \in &\Big\lbrace 0, \pm 1, \pm \frac{i}{\sqrt{3}}  \Big\rbrace \, . \label{eq:b_indices}
\end{align}
The transcendental constants that appear in these integrals just include $\log(2)$ and $\zeta_n$.

\subsection{Families 2 and 5}

Families 2 and 5 are the most complicated, as they feature the additional root
\begin{align}
\sqrt{2 M^4 +2 M^2 (s-4)+s(s-2) \pm s (2 M^2+s) \sqrt{-s}\sqrt{4-s\,}} \, ,
\end{align}
(where we have now set $m_t =1$ for simplicity) and contain the elliptic sunrise integral with two distinct masses as a subtopology. Therefore, it is clear that we can no longer express these integrals in terms of traditional multiple polylogarithms. Nevertheless, it remains possible to find an $\epsilon$-factorized form of the corresponding differential equations by following the steps outlined in~\cite{Gorges:2023zgv}. In the following discussion, we aim to provide only the essential information needed to write down the integrals (included in the ancillary files) that result in an $\epsilon$-factorized differential equation for the elliptic families 2 and 5. Only three new transcendental functions are needed to define these integrals, as we will see below. For a comprehensive understanding of how to systematically derive these integrals, we refer readers to~\cite{Gorges:2023zgv}.

Families 2 and 5 contain only a single elliptic sector, which is elliptic on its maximal cut and contains three master integrals. This specific sector corresponds to the elliptic two-mass sunrise, which has been intensively studied in Section 3.3 of~\cite{Gorges:2023zgv}. To bring this system into canonical form, we particularly need the elliptic period $\varpi_0$, which can also be expressed through the standard elliptic integrals in the following way 
\begin{align}
\varpi_0 = \frac{2i}{\pi}\frac{ \sqrt{-M^2 \left(M^2-4\right)} \, K\left(\frac{1}{2} + \frac{M^4-2 (s+2) M^2+(s-4) s}{2 \left(M^2-s\right)  \sqrt{\big(s-(M+2)^2\big)\big(s-(M-2)^2\big)}}\right)}{\pi  \, \sqrt{s-M^2} \, \sqrt[4]{\big(s-(M+2)^2\big)\big(s-(M-2)^2\big)}} \,,
\label{eq:period}
\end{align}
where 
\begin{align}
K(z) = \int_0^1 \frac{1}{\sqrt{\left(1-t^2\right) \left(1-z \, t^2\right)}} \, \text{d}t
\end{align}
is the complete elliptic integral of the first kind. This new transcendental function can be seen as the leading singularity of the two-mass sunrise integral with unit propagator powers and thus generalizes the square-root leading singularity of the non-elliptic sunrise integral. From this perspective, it is natural that this function is needed in defining a canonical basis. Moreover, as extensively discussed in~\cite{Gorges:2023zgv}, another new transcendental function is needed to $\epsilon$-factorize the two-mass sunrise. This second new function is defined as a one-fold iterated integral over $\varpi_0$, given by
\begin{equation}
	G_1 = 2 \int_0^s \frac{(M^2-1)(M^2-s'-4)}{(M^2+3s'-4)^2}\, \varpi_0(s')\, \mathrm ds' \, .
\label{eq:g1}
\end{equation}
This integral is related to an elliptic integral of the third kind (see~\cite{Gorges:2023zgv}) and is connected to a new singularity structure (puncture) on the associated elliptic curve. For family 5, these are all the transcendental functions that are needed to bring the system into canonical form. In the case of family 2, we need a second one-fold iterated integral over $\varpi_0$, namely 
\begin{equation}
	G_2 = 2\sqrt{M^2}\sqrt{4-M^2} \int_0^s \tfrac{ \left(M^4+3 M^2 s'^2-13 M^2 s'-4 M^2-3 s'^3+16 s'\right)}{(s'-4)^2 s' \left(M^2+3 s'-4\right)^2}\sqrt{-s'}\sqrt{4-s'} \varpi_0(s')  \, \mathrm ds' \, .
\label{eq:g2}
\end{equation}
Also, this integral can be related to a new singularity structure, which---unlike the previous case---arises from a mixing between polylogarithmic and elliptic sectors, rather than from the maximal cuts of a single sector. With $\varpi_0, G_1$, and $G_2$, family 2 can also be brought into canonical form as they appear in the rotations from the initial Laporta basis to the canonical one. The precise definition of all canonical integrals using the Laporta integrals given in \ref{sec:basis_description} and the three transcendental functions from above are listed in the ancillary files.

Once we have put the differential equations that describe these families in canonical form, we can express each of the integrals they contain as iterated integrals over a bespoke set of integration kernels
\begin{align}
\mathbb{G}(k_{i_1},k_{i_2},\dots)(s) = \int_0^s k_{i_1}(s') \, \mathbb{G}(k_{i_2},\dots)(s') \, \text{d}s' \,,
\end{align}
where the integration kernels $k_j(s)$ will be given below. The algebraic roots that appear in these integration kernels are given by
\begin{align}
r_1 &= \sqrt{-s}\,, ~~~~ r_2 = \sqrt{4-s}\,, ~~~~ r_3 = \sqrt{M^2}\,, ~~~~ r_4 = \sqrt{4-M^2}\,, \nonumber \\ \label{eq:roots} 
r_5 &= \sqrt{M^4-2 M^2 s-2 M^2+s^2-2 s+1}\,, \\ \nonumber
r_6 &= \sqrt{2 M^4 +2 M^2 (s-4)+s(s-2) - s (2 M^2+s) \sqrt{-s}\sqrt{4-s\,}}\,, \\ \nonumber
r_7 &= \sqrt{2 M^4 +2 M^2 (s-4)+s(s-2) + s (2 M^2+s) \sqrt{-s}\sqrt{4-s\,}} \,.
\end{align} 
We choose not to rationalize any of these roots, as this allows us to keep the number of integration kernels (and hence the size of the expressions) to a minimum. 

The integration kernels can be split into those that depend on the elliptic curve associated with the two-mass sunrise integral and those that do not. We label the former by $k^\text{E}_i$, and the latter by $k_j$. The elliptic dependence appears with the elliptic period $\varpi_0$ defined in \eqref{eq:period} and the one-fold iterated integrals $G_1, G_2$ given in \eqref{eq:g1} and \eqref{eq:g2}, respectively. To simplify the expressions that follow, we write $G_1, G_2$ in our notation of iterated integrals over independent one-forms as
\begin{align}
G_1 &= \frac{8}{3}(M^2-4)(M^2-1)\mathbb{G}(k^\text{E}_1) - \frac{2}{9} (M^2-1) \mathbb{G}(k^\text{E}_3)\, , \\
G_2 &= 2 r_3 r_4 \mathbb{G}(k^\text{E}_8)  \,.
\end{align}
The full set independent integration kernels is then given by
\begin{align}\label{eq:kersBeginning}
k_1 &= \frac{1}{s}\,, ~~~~ k_2 = \frac{1}{s-4}\,, ~~~~~ k_3 = \frac{1}{s-M^2}\,, ~~~~ k_4 = \frac{1}{M^2+s-4}\,, ~~~~ k_5 = \frac{1}{M^2+3 s-4}\,, \\ 
k_6 &= \frac{M^2}{M^2 s+\left(M^2-1\right)^2}\,, ~~~ k_7 = \frac{M^2-s+4}{-2 \left(M^2+4\right) s+\left(M^2-4\right)^2+s^2}\,, \\ 
k_8 &= \frac{M^2-s+1}{-2 \left(M^2+1\right) s+\left(M^2-1\right)^2+s^2}\,, \\ 
k_9 &= \frac{\left(M^2-1\right)^2 s+\left(M^4-3 M^2+4\right) M^2}{\left(M^2-1\right)^2 s^2+2 \left(M^4-3 M^2+4\right) M^2 s+\left(M^2-4\right)^2 M^4}\,, \\ 
k_{10} &= \frac{r_1 r_2}{(s-4) s}\,, ~~~~~~~ k_{11} = \frac{r_1 r_2 \left(M^2 s-2 M^2+2\right)}{(s-4) s \left(M^4+M^2 s-2 M^2+1\right)}\,, \\ 
k_{12} &= \frac{r_1 r_2 \left(-M^8+4 M^6+M^4 s^2-4 M^4 s-2 M^2 s^2+12 M^2 s+s^2\right)}{(s-4) s \left(M^8+2 M^6 (s-4)+M^4 \left(s^2-6 s+16\right)-2 M^2 (s-4) s+s^2\right)}\,, \\ 
k_{13} &= \frac{r_1 r_2}{(s-4) s \left(M^2+s-4\right)}\,, ~~~~~~ k_{14} = \frac{r_1 r_2}{(s-4) s \left(M^2+3 s-4\right)}\,, \\ 
k_{15} &= \frac{r_5}{M^4-2 M^2 (s+1)+(s-1)^2}\,, ~~~~~~ k_{16} = \frac{\left(M^2-1\right) r_5}{s \left(M^4-2 M^2 (s+1)+(s-1)^2\right)}\,, \\ 
k_{17} &= \frac{r_1 r_2 r_5 \left(M^2-3 s+3\right)}{4 (s-4) s \left(M^4-2 M^2 (s+1)+(s-1)^2\right)}\,, \\ 
k_{18} &= \frac{\left(r_6+r_7\right) \left(3 M^4+M^2 (s-4)-s\right)}{M^8+2 M^6 (s-4)+M^4 \left(s^2-6 s+16\right)-2 M^2 (s-4) s+s^2} \nonumber \\
&+\frac{r_1 r_2 \left(r_6-r_7\right) \left(2 M^6+M^4 (s-8)+M^2 (s-2) s-s^2\right)}{(s-4) s \left(M^8+2 M^6 (s-4)+M^4 \left(s^2-6 s+16\right)-2 M^2 (s-4) s+s^2\right)}\,, \\ 
k_{19} &= \frac{\left(r_6-r_7\right) \left(3 M^4+M^2 (s-4)-s\right)}{M^8+2 M^6 (s-4)+M^4 \left(s^2-6 s+16\right)-2 M^2 (s-4) s+s^2} \nonumber \\
&+\frac{r_1 r_2 \left(r_6+r_7\right) \left(2 M^6+M^4 (s-8)+M^2 (s-2) s-s^2\right)}{(s-4) s \left(M^8+2 M^6 (s-4)+M^4 \left(s^2-6 s+16\right)-2 M^2 (s-4) s+s^2\right)}\,, \\ 
k_{20} &= \frac{\left(r_6-r_7\right) \left(M^2 (s+2)-s\right)}{M^8+2 M^6 (s-4)+M^4 \left(s^2-6 s+16\right)-2 M^2 (s-4) s+s^2} \nonumber \\
&+\frac{r_1 r_2 \left(r_6+r_7\right) \left(2 M^6+2 M^4 (2 s-7)+M^2 \left(s^2-8 s+24\right)-(s-6) s\right)}{(s-4) s \left(M^8+2 M^6 (s-4)+M^4 \left(s^2-6 s+16\right)-2 M^2 (s-4) s+s^2\right)}\,, \\ 
k_{21} &= \frac{\left(r_6+r_7\right) \left(M^2 (s+2)-s\right)}{M^8+2 M^6 (s-4)+M^4 \left(s^2-6 s+16\right)-2 M^2 (s-4) s+s^2} \nonumber \\
&+\frac{r_1 r_2 \left(r_6-r_7\right) \left(2 M^6+2 M^4 (2 s-7)+M^2 \left(s^2-8 s+24\right)-(s-6) s\right)}{(s-4) s \left(M^8+2 M^6 (s-4)+M^4 \left(s^2-6 s+16\right)-2 M^2 (s-4) s+s^2\right)}\,, \\ 
k_{22} &= \frac{r_1 r_2 r_3 r_4 \left(r_6+r_7\right) \left(M^4+s\right)}{(s-4) s \left(M^8+2 M^6 (s-4)+M^4 \left(s^2-6 s+16\right)-2 M^2 (s-4) s+s^2\right)} \nonumber \\
&-\frac{r_3 r_4 \left(r_6-r_7\right) \left(M^4-4 M^2-s\right)}{s \left(M^8+2 M^6 (s-4)+M^4 \left(s^2-6 s+16\right)-2 M^2 (s-4) s+s^2\right)}\,, \\ 
k_{23} &= \frac{r_1 r_2 r_3 r_4 \left(r_6-r_7\right) \left(M^4+s\right)}{(s-4) s \left(M^8+2 M^6 (s-4)+M^4 \left(s^2-6 s+16\right)-2 M^2 (s-4) s+s^2\right)} \nonumber \\
&-\frac{r_3 r_4 \left(r_6+r_7\right) \left(M^4-4 M^2-s\right)}{s \left(M^8+2 M^6 (s-4)+M^4 \left(s^2-6 s+16\right)-2 M^2 (s-4) s+s^2\right)}\,, \\ 
k_{24} &= \frac{r_6 r_7 \left(M^4 (s-2)-2 M^2 (s-4)+2 s\right)}{(s-4) s \left(M^8+2 M^6 (s-4)+M^4 \left(s^2-6 s+16\right)-2 M^2 (s-4) s+s^2\right)}
\end{align}
\begin{align}
k_0^{\text{E}} &= \frac{M^2+3 s-4}{s \left(M^2-s\right) \left(M^4-2 M^2 s-8 M^2+s^2-8 s+16\right)\, \varpi_0^2 }\,, ~~~~ k_1^{\text{E}} = \frac{\varpi_0}{\left(M^2+3 s-4\right)^2}\,, \\ 
k_2^{\text{E}} &= \varpi_0\,, ~~~~ k_3^{\text{E}} = \frac{3 \varpi_0}{M^2+3 s-4}\,,  ~~~~ k_4^{\text{E}} = \frac{\varpi_0}{M^2+s-4}\,, ~~~~ k_5^{\text{E}} = \frac{\varpi_0}{s-4}\,, ~~~~ k_6^{\text{E}} = \frac{\varpi_0}{s}\,, \\ 
k_7^{\text{E}} &= \frac{r_1 r_2 \varpi_0 \left(-M^4+3 M^2 s+4 M^2+2 s^2-8 s\right)}{(s-4) s \left(M^2+3 s-4\right)}\,, \\ 
k_8^{\text{E}} &= \frac{r_1 r_2 \varpi_0 \left(M^4+3 M^2 s^2-13 M^2 s-4 M^2-3 s^3+16 s\right)}{(s-4)^2 s \left(M^2+3 s-4\right)^2}\,, \\ 
k_9^{\text{E}} &= \varpi_0 \Big(-\frac{16 M^2}{M^2-s}+\frac{\left(M^2-4\right) M^2}{s}+\frac{16 \left(M^2-4\right) \left(4 M^2+3\right)}{3 \left(M^2+3 s-4\right)} -\frac{22}{9} \left(M^2+2\right) \\ \nonumber
&~~ -\frac{64 \left(M^2-4\right) \left(M^2-1\right) \left(11 M^2-8\right)}{9 \left(M^2+3 s-4\right)^2}+\frac{256 \left(M^2-4\right)^2 \left(M^2-1\right)^2}{3 \left(M^2+3 s-4\right)^3} \\ \nonumber
&~~ +\frac{64 M^2 \left(M^2-s-4\right)}{M^4-2 M^2 s-8 M^2+s^2-8 s+16}-\frac{s}{3}\Big)\,, \\ 
k_{10}^{\text{E}} &= \frac{G_1}{s}\,, ~~~~ k_{11}^{\text{E}} = \frac{G_1}{s-M^2}\,, ~~~~ k_{12}^{\text{E}} = \frac{G_1}{M^2+s-4}\,, ~~~~ k_{13}^{\text{E}} = \frac{G_1}{M^2+3 s-4}\,, \\ 
k_{14}^{\text{E}} &= \frac{G_1 \left(M^2-s+4\right)}{M^4-2 M^2 s-8 M^2+s^2-8 s+16}\,,  ~~~~ k_{15}^{\text{E}} = \frac{G_1 r_1 r_2}{(s-4) s}\,, \\ 
k_{16}^{\text{E}} &= \frac{G_1 r_1 r_2}{(s-4) s \left(M^2+3 s-4\right)}\,,  ~~~~ k_{17}^{\text{E}} = \frac{G_1 \varpi_0 \left(M^2-s-4\right)}{\left(M^2+3 s-4\right)^2}\,, \\ 
k_{18}^{\text{E}} &= \frac{G_1 \left(M^2+3 s-4\right)}{s \left(M^2-s\right) \left(M^4-2 M^2 s-8 M^2+s^2-8 s+16\right)\,\varpi_0^2 }\,, \\ 
k_{19}^{\text{E}} &= \frac{G_2 r_1 r_2}{(s-4) s}\,, ~~~~ k_{20}^{\text{E}} = \frac{G_2 r_1 r_2}{(s-4) s \left(M^2+s-4\right)}\,, \\ 
k_{21}^{\text{E}} &= G_2 \Big(\frac{1}{M^2-s}+\frac{3}{M^2+3 s-4}+\frac{2 \left(M^2-s+4\right)}{M^4-2 M^2 (s+4)+(s-4)^2}+\frac{1}{s-4}+\frac{1}{2 s}\Big), \\ 
k_{22}^{\text{E}} &= \frac{G_2 \left(M^2+3 s-4\right)}{s \left(M^2-s\right) \left(M^4-2 M^2 (s+4)+(s-4)^2\right)\,\varpi_0^2}\,, \\ 
k_{23}^{\text{E}} &= G_1^2 \Big(\frac{2}{s-M^2}-\frac{18}{M^2+3 s-4}-\frac{4 \left(M^2-s+4\right)}{M^4-2 M^2 (s+4)+(s-4)^2}+\frac{1}{s}\Big)\,, \\ 
k_{24}^{\text{E}} &= \frac{G_1^2 \left(M^2+3 s-4\right)}{s \left(M^2-s\right) \left(M^4-2 M^2 s-8 M^2+s^2-8 s+16\right)\,\varpi_0^2}\,, \\ 
k_{25}^{\text{E}} &= \frac{G_1 G_2 \left(M^2+3 s-4\right)}{s \left(s-M^2\right) \left(M^4-2 M^2 (s+4)+(s-4)^2\right)\,\varpi_0^2}\,, \\ 
k_{26}^{\text{E}} &= \frac{G_1^3 \left(M^2+3 s-4\right)}{s \left(M^2-s\right) \left(M^4-2 M^2 s-8 M^2+s^2-8 s+16\right)\,\varpi_0^2}\,, \\ 
k_{27}^{\text{E}} &= \frac{G_1^2 G_2 \left(M^2+3 s-4\right)}{s \left(M^2-s\right) \left(M^4-2 M^2 (s+4)+(s-4)^2\right)\,\varpi_0^2}\,, \\ 
k_{28}^{\text{E}} &= \frac{G_1^4 \left(M^2+3 s-4\right)}{s \left(M^2-s\right) \left(M^4-2 M^2 s-8 M^2+s^2-8 s+16\right)\,\varpi_0^2} \, .
\label{eq:kersEnd}
\end{align}
We include the definitions of these kernels alongside our iterated integral representations of the master integrals in families 2 and 5 in the ancillary files. Note that some kernels have a singularity for $  s\rightarrow M^2$. These singularities cancel out in each master integral, since none of our diagrams have a cut in the physical region involving a single massive line of mass $M^2$ (see appendix \ref{sec:basis_description}).

\subsection{Cross-checks}
\label{sec:numerics}

As a cross-check, we have compared our iterated integral results for the master integrals against a numeric evaluation using AMFlow~\cite{Liu:2022chg}, for instance, at $M^2 = 9/10$ and $s = \pm 1/10$ or $M^2 = 1/3$ and $s = \pm 1/75$ depending on their radius of convergence. The polylogarithmic master integrals have been checked numerically up to 40 digits through weight five and to 20 digits at weight six using GiNaC~\cite{Bauer:2000cp}, while the elliptic master integrals have been checked numerically up to 13 digits through weight six by expanding the integration kernels in $s$ before integrating them analytically (and subsequently evaluating those using established tools). We have also checked that the integrals can be evaluated in the physical region by expanding around $s=0$ (first assuming $s<0$) and then analytically continuing them to $s>0$. Since all elliptic integrals are regular for $s=0$, the analytic continuation from $s<0$ to $s>0$ is trivial, and one can use the series expansions in both regions. These numerical checks complement our analytic confirmation that our expressions exhibit the correct behavior at $s = 0$ and fulfil the canonical differential equation through weight six. Note that a numeric evaluation of the elliptic integrals for large values of s is possible as well by calculating a series expansion around the threshold and analytically continuing the appearing logarithms and roots appropriately. There is no conceptual difficulty, as we are using generalized series expansions for the numeric evaluation.

Note that, due to electro-weak symmetry breaking, the Yukawa sector is coupled to the electro-weak sector through the Higgs vacuum expectation value $v$, as we have 
\begin{align}
y_t \propto \frac{m_t}{v} \propto \sqrt{\alpha_\text{EW}} \frac{m_t}{\sin \theta_W \, m_W} \,.
\end{align}
We can see that the Yukawa coupling can be expressed in terms of the electro-weak gauge boson masses as well as the electro-weak coupling and the vacuum expectation value. As a result, the renormalisation of all of these parameters is not independent. Therefore, one must consider diagrams where the Higgs boson couples to the electro-weak gauge bosons when calculating scattering amplitudes featuring massive electro-weak gauge bosons and can not consider only diagrams where the Higgs couples exclusively to the top quark~\cite{Denner:1991kt,Denner:2019vbn}.
In order to construct a gauge-independent amplitude, we also need to incorporate contributions from diagrams in which the Higgs boson is emitted from electro-weak gauge bosons or their Goldstone bosons. 
One can check that the missing two-loop diagrams where the Higgs couples to $W$ bosons are purely polylogarithmic. Therefore, and any elliptic dependence on the gauge parameter $\xi_W$ must vanish independently when considering only the diagrams calculated in this paper. We have checked that this is indeed the case.

\subsection{Analytic Properties}
\label{sec:analytic_properties}

It is interesting to analyze the analytic properties that are exhibited by these families of integrals, given the significant progress that has recently been made in our understanding of the singularities and discontinuities of Feynman integrals~\cite{Caron-Huot:2016owq,Drummond:2017ssj,Bourjaily:2020wvq,Hannesdottir:2022xki,Fevola:2023kaw,Helmer:2024wax,Caron-Huot:2024brh,Correia:2025yao}. Consider, for instance, the integrals that appear in families 1 and 4; the only symbol letter that appears in the first entry of these integrals is
\begin{equation}
\log\left(\frac{1-x}{1+x}\right) = \log \left( \frac{\sqrt{4m_t^2 - s} - \sqrt{-s}}{\sqrt{4m_t^2 - s} + \sqrt{-s}} \right) \, ,
\end{equation}
matching our na\"ive expectation that the first threshold and pseudo-threshold discontinuities should be algebraic, while the $m_t^2 \to 0$ singularity is logarithmic (this follows from counting the number of on-shell propagators in the bubble and tadpole Landau diagrams, which are naturally associated with these singularities~\cite{Landau:1959fi, Hannesdottir:2021kpd}).\footnote{While one might worry that the three-particle threshold in family 4 (in which one massless and two massive particles are cut) would give rise to a logarithmic singularity, it can be shown that this singularity is also algebraic by carrying out the appropriate blowups~\cite{Hannesdottir:2024hke,to_appear_landau_bootstrap}.} We also see that the integrals in these families respect a Galois symmetry at the symbol level with respect to flipping the sign in front of either $\sqrt{-s}$ or $\sqrt{4m_t^2-s}$, once the algebraic prefactors are taken into account.

Additional types of structure can be observed in the integrals of family 3, especially when their polylogarithmic symbol is expressed in terms of $s$, $m_t$, and $M$. In terms of these variables, twelve $d\log$ kernels appear and are given by
\begin{gather}
L_1 = d\log(M^2)\,, \qquad L_2 = d\log(m_t^2)\,, \qquad L_3 = d\log(s)\, , \qquad L_4 = d\log(s - 4 m_t^2)\, , \nonumber \\
L_5 = d\log(M^2 - m_t^2)\, , \qquad L_6 = d\log(M^4 - 2 M^2 m_t^2 + m_t^4 + M^2 s)\, , \nonumber \\ 
L_7 = d\log(M^4 - 2 M^2 m_t^2 + m_t^4 - 2 M^2 s - 2 m_t^2 s + s^2) \, ,  \label{eq:family_three_momentum_alphabet} \\
L_8 = d\log\bigg(\frac{-2 M^2 + 2 m_t^2 - s - r_1 r_2}{-2 M^2 + 2 m_t^2 - s + r_1 r_2} \bigg) \, , \qquad
L_9 = d\log\bigg(\frac{2 m_t^2 - s - r_1 r_2}{2 m_t^2 - s + r_1 r_2} \bigg) \, , \nonumber \\
L_{10} = d\log \bigg(\frac{M^2 + m_t^2 - s - r_5}{M^2 + m_t^2 - s + r_5} \bigg) \, , \qquad 
L_{11} = d\log \bigg(\frac{M^2 - m_t^2 + s - r_5}{M^2 - m_t^2 + s + r_5} \bigg)\, , \nonumber \\
L_{12} = d\log \bigg(\frac{M^2 s + 3 m_t^2 s - s^2 - r_1 r_2 r_5}{M^2 s + 3 m_t^2 s - s^2 + r_1 r_2 r_5}\bigg) \, , \nonumber
\end{gather} 
where $r_1$, $r_2$, and $r_5$ were defined in~\eqref{eq:roots} (recall that in those definitions, $m_t$ was set to 1). This already represents a massive simplification compared to the sixty-nine integration kernels that appear in the $G_{\vec{a}}(z)$ and $G_{\vec{b}}(x)$ functions that draw upon the indices~\eqref{eq:a_indices} and~\eqref{eq:b_indices}.\footnote{This number can be checked by computing the symbols of the expressions for the master integrals provided in the ancillary files. In fact, the alphabet in~\eqref{eq:family_three_momentum_alphabet} is precisely the set of letters predicted by \texttt{SOFIA} and \texttt{Effortless} \cite{Correia:2025yao, Effortlessxxx}, as long as one tells the code that $r_1 r_2$ appears as a double root.} In this basis of kernels, only $L_1$, $L_2$, $L_3$, $L_9$, and $L_{10}$ appear in the first entry (although $L_{11}$ might also have been expected to appear, since it only encodes the same algebraic and logarithmic singularities that appear in the other first-entry letters). Each integral is also Galois even with respect to $r_5$, since this singularity cannot appear on the physical Riemann sheet (recall that this symmetry also takes algebraic prefactors into account); some of the master integrals in this family are also Galois even with respect to $r_1$, due to the absence of a massless threshold. Moreover, while the integrals in this family do not have sufficiently complex kinematic dependence to obey any genealogical constraints~\cite{Hannesdottir:2024cnn} or extended Steinmann relations~\cite{Caron-Huot:2019bsq}, they can be observed to respect the adjacency restriction that the pairs
\begin{equation}
\{L_4, L_7\}\, , \, \{L_6, L_7\}\, , \, \{L_6, L_{12}\}\, , \, \{L_7, L_8\}\, , \, \{L_7, L_9\}\, , \, \{L_8, L_{10}\}\, , \,  \{L_8, L_{11}\}\, , \,  \{L_{12}, L_{12}\}
\end{equation}
never appear next to each other when expressed in terms of iterated integrals. 

We might next ask how restrictive all these properties are, when we put them all together. For instance, let us focus on the integral $I^{(3)}_{1, 1, 0, 1, 1, 1, 0}$, which has no massless threshold and therefore cannot have a discontinuity that starts at $s=0$ on the physical Riemann sheet. This implies that the only $d\log$ kernels that can appear in the first entry are $L_1$, $L_2$, $L_{9}$, $L_{10}$, and $L_3 + L_{11}$. It also implies that the integral must be Galois even with respect to $r_1$, as well as with respect to $r_5$. Putting these constraints together---along with the empirical adjacency constraints listed above---we find the following number of symbols can appear at weight four:
\begin{center}
\begin{tabular} {c | c}
constraint & number of weight four symbols \\ \hline
integrability & 4114 \\
first entry in $\{L_1, L_2, L_{9}, L_{10}, L_3 + L_{11}\}$ & 790 \\
Galois even in $r_1$ and $r_5$ & 233 \\
empirical adjacency constraints & 136 \\
\end{tabular}
\end{center}
Thus, we see that these properties by themselves are not yet sufficient to identify this master integral---which is not surprising, given that this is not the only integral in family 3 that satisfies all these constraints. It would be interesting to derive further constraints in order to bootstrap the symbol of $I^{(3)}_{1, 1, 0, 1, 1, 1, 0}$ directly, for instance following the strategy outlined in~\cite{Hannesdottir:2024hke}. Even more ambitiously,  the same strategy should prove powerful enough to bootstrap the elliptic integrals in sectors 2 and 5. However, we leave these investigations to future work.

\section{Conclusions}
\label{sec:conclusion}

In this paper, we have presented analytic results for all of the two-loop integrals that appear at $\mathcal{O}(g^3 g_s^2)$ in which the Higgs boson couples to the top quark, keeping full dependence on the masses of the Higgs boson, top quark, and electroweak bosons. Working in $d = 4 - 2 \epsilon$ dimensions, we have calculated these integrals through $\mathcal{O}(\epsilon^2)$ by first solving the system of differential equations for five families of master integrals and then rotating the resulting solutions into $\epsilon$-factorized form. The boundary conditions for these differential equations were then determined in the $s \to 0$ limit by imposing the absence of pseudo-threshold singularities (whenever this singularity can be distinguished from the threshold), or using the method of regions. The ancillary files that contain analytic expressions for the master integrals, as well as the rotations that define the canonical master integrals, can be found on the arXiv and on Zenodo~\cite{marzucca_2025_14843619}.

Three of the families of integrals we consider can be evaluated in terms of iterated integrals that involve only $d\log$ integration kernels. We have presented the results for these integrals in terms of multiple polylogarithms in a form that can be symbolically manipulated and numerically evaluated using public codes~\cite{Bauer:2000cp, Vollinga:2004sn, Panzer:2014caa, Duhr:2019tlz}. The remaining two families give rise to integration kernels that involve the elliptic curve associated with the two-mass sunrise integral. We have presented the results for these integrals in terms of iterated integrals over the kernels shown in equations~\eqref{eq:kersBeginning}-\eqref{eq:kersEnd}. Since only a single elliptic curve appears in each integral, we believe it should be possible to express these integrals in terms of elliptic multiple polylogarithms (as defined in~\cite{Brown:2011wfj, Broedel:2017kkb}). However, we have left this possibility to future work. 

Clearly, the next step will be to assemble the $gg \to H$ amplitude at $\mathcal{O}(g^3 g_s^2)$~\cite{ggH_amplitude_paper}. This will require computing additional integrals, in which the Higgs does not couple to the top quark. While numerical results for the Standard Model cross-section have long been available~\cite{Actis:2008ug}, arriving at an iterated integral expression for this amplitude will give us better numerical control over Standard Model predictions and allow us to better estimate the theoretical uncertainties associated with the masses of heavy Standard Model particles. It will also allow us to search for new and unexpected types of mathematical structure in the $gg \to H$ amplitude, which could, in turn, be leveraged to help compute mixed strong-electroweak corrections to this process at higher perturbative orders. 

In fact, by explicitly computing the two-loop integrals considered in this paper through $\mathcal{O}(\epsilon^2)$, we have already begun to lay the groundwork for determining mixed strong-electro\-weak corrections to this process at higher orders. Clearly, the higher-loop integrals that contribute to $gg \to H$ will also involve integrals over the two-mass sunrise elliptic curve, but it will be interesting to see which other geometries will appear, such as new elliptic curves or higher-dimensional manifolds. Identifying what special functions appear at higher orders remains an interesting question. The answer will dictate what types of technology need to be developed to get Standard Model corrections to $gg \to H$ under both analytic and numerical control.

\section*{Acknowledgements}
\label{sec:acknowledgements}

We thank Thomas Gehrmann, Kay Schönwald, Nikolaos Syrrakos, Lorenzo Tancredi, and Fabian Wagner for their continued and helpful discussions. This work has been supported by the European Union’s Horizon 2020 research and innovation program EWMassHiggs (Marie Sk{\l}odowska Curie Grant agreement ID: 101027658) and by the European Research Council (ERC) under the European Union’s research and innovation programme grant agreement 101019620 (ERC Advanced Grant TOPUP). AJM is supported by the Royal Society grant URF{\textbackslash}R1{\textbackslash}221233, CN was supported in part by the Excellence Cluster ORIGINS funded by the Deutsche Forschungsgemeinschaft (DFG, German Research Foundation) under Germany’s Excellence Strategy – EXC-2094-390783311 and in part by the European Research Council (ERC) under the European Union’s research and innovation program grant agreements 949279 (ERC Starting Grant HighPHun).

\appendix

\section{Master Integral Basis}
\label{sec:basis_description}

In this appendix, we present the details of our initial basis of master integrals before it has been rotated into canonical form. This makes it easy to specify each basis element in terms of the powers of the propagators specified in equations~\eqref{eq:props_1} through~\eqref{eq:props_5}. We reprise these lists of propagators below and include the matrices that rotate this basis into $\epsilon$-factorized form in ancillary files. 

\subsection{Family 1}
\label{subsec:F1}

The propagators that define this family are given by
\begin{align}
D^{(1)}_i &= \{\, k_1^2 - m_t^2\, ,\, (k_1 + p_1)^2 - m_t^2\, ,\, (k_1 - p_2)^2 - m_t^2\, ,\, \nonumber \\ 
&\qquad(k_1 + k_2)^2\, ,\, (k_2 - p_1)^2 - m_t^2\, ,\, (k_2 + p_2)^2 - m_t^2\, ,\, k_2^2 \,\} \, , \nonumber
\end{align}
and it is spanned by ten integrals:
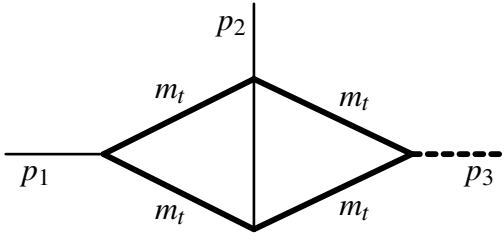
\begin{center}
\begin{longtable}{p{8cm} | p{8cm}}
\begin{center} \begin{tikzpicture}[baseline=(current bounding box.center), line width=2.2, scale=1,line cap=round]
    \draw[black] (-2,0) -- node[above,xshift=-.1cm] {$m_t$} (0,1);
    \draw[black] (2.1,0) -- node[below,xshift=0.3cm,yshift=.05cm] {$m_t$} (0,-1);
    \draw[black] (0,-1) -- node[below,xshift=-.1cm] {$m_t$} (-2,0);
    \draw[black] (0,1) to node[above,xshift=0.3cm,yshift=-.1cm]{$m_t$} (2.1,0);
    \draw[black, line width=1] (0,1) -- (0,-1);
    \draw[black, line width=1] (0,1) -- (0,2);
    \draw[black, line width=1] (-2,0) -- (-3.3,0);
    \draw[black, dashed] (2.1,0) -- (3.3,0);
	\node [black] at (-2.9,-.3) {$p_1$};
	\node [black] at (-.3,1.7) {$p_2$};
	\node [black] at (3,-.3) {$p_3$};
\end{tikzpicture} \end{center}
&
\begin{center} \begin{tikzpicture}[baseline=(current bounding box.center), line width=2.2, scale=1,line cap=round]
    \draw[black] (2.1,0) -- node[below,xshift=0.3cm,yshift=.05cm] {$m_t$} (0,-1);
    \draw[black] (0,1) to node[above,xshift=0.3cm,yshift=-.1cm]{$m_t$} (2.1,0);
    \draw[black, line width=1] (0,1) to [bend left] (0,-1);
    \draw[black] (0,1) to [bend right] node[left] {$m_t$} (0,-1);
    \draw[black, line width=1] (0,-1) -- (-.7,-1.85);
    \draw[black, line width=1] (0,1) -- (-.7,1.85);
    \draw[black,dashed] (2.1,0) -- (3.3,0);
	\node [black] at (-1,1.6) {$p_2$};
	\node [black] at (-1,-1.6) {$p_1$};
	\node [black] at (3,-.3) {$p_3$};
\end{tikzpicture} \end{center}
\\*[-.8cm]
\begin{center} $I^{(1)}_{1, 1, 0, 1, 1, 2, 0}$\ , \ $I^{(1)}_{1, 1, 0, 1, 1, 1, 0}$ \end{center} &
\begin{center} $I^{(1)}_{2, 0, 0, 1, 2, 1, 0}$\ , \ $I^{(1)}_{1, 0, 0, 2, 1, 1, 0}$\ , \ $I^{(1)}_{2, 0, 0, 1, 1, 1, 0}$ \end{center} \\ \hline
\begin{center} \begin{tikzpicture}[baseline=(current bounding box.center), line width=2.2, scale=1,line cap=round]
    \draw[black] (2.1,0) -- node[below,xshift=0.3cm,yshift=.05cm] {$m_t$} (0,-1);
    \draw[black] (0,1) to node[above,xshift=0.3cm,yshift=-.1cm]{$m_t$} (2.1,0);
    \draw[black, line width=1] (0,1) to [bend left] (0,-1);
    \draw[black] (0,1) to [bend right] node[left] {$m_t$} (0,-1);
    \draw[black, dashed] (0,-1) -- (-.7,-1.85);
    \draw[black, line width=1] (0,1) -- (-.7,1.85);
    \draw[black, line width=1] (2.1,0) -- (3.3,0);
	\node [black] at (-1,1.6) {$p_1$};
	\node [black] at (-1,-1.6) {$p_3$};
	\node [black] at (3,-.3) {$p_2$};
\end{tikzpicture} \end{center}
&
\begin{center} \begin{tikzpicture}[baseline=(current bounding box.center), line width=2.2, scale=1,line cap=round]
    \draw[black, line width=1] (2,0) to (0,0);
    \draw[black] (0,0) to [bend left=45*-1] node[below,yshift=-.05cm]{$m_t$} (2,0);
    \draw[black] (0,0) to [bend right=45*-1] node[above,yshift=-.05cm]{$m_t$} (2,0);
    \draw[black, line width=1] (0,0) -- (-1,-.85);
    \draw[black, line width=1] (0,0) -- (-1,.85);
    \draw[black,dashed] (2.1,0) -- (3.3,0);
	\node [black] at (-0.6,-1) {$p_1$};
	\node [black] at (-0.6,1) {$p_2$};
	\node [black] at (3,-.3) {$p_3$};
\end{tikzpicture} \end{center}
\\*[-.8cm]
\begin{center} $I^{(1)}_{1, 0, 1, 1, 2, 0, 0}$ \end{center} &
\begin{center} $I^{(1)}_{0, 0, 2, 2, 1, 0, 0}$\ , \ $I^{(1)}_{0, 0, 2, 1, 2, 0, 0}$ \end{center} \\ \hline
\begin{center} \begin{tikzpicture}[baseline=(current bounding box.center), line width=2.2, scale=1,line cap=round]
    \draw[black] (2,0) to [bend left] node[below] {$m_t$} (0,0);
    \draw[black] (0,0) to [bend left] node[above,yshift=-.05cm]{$m_t$} (2,0);
    \draw[black] (2,0) to [bend right=45*-1] (2.4,0.4);
    \draw[black] (2.4,0.4) to [bend right=45*-1] (2.8,0);
    \draw[black] (2.8,0) to [bend right=45*-1] (2.4,-0.4);
    \draw[black] (2.4,-0.4) to [bend right=45*-1] (2,0);
    \draw[black, line width=1] (0,0) -- (-1,-.85);
    \draw[black, line width=1] (0,0) -- (-1,.85);
    \draw[black,dashed] (1.8,-1.1) -- (1.94,-.1);
	\node [black] at (-0.6,-1) {$p_1$};
	\node [black] at (-0.6,1) {$p_2$};
	\node [black] at (2.2,-.9) {$p_3$};
	\node [black] at (3.13,0) {$m_t$};
\end{tikzpicture} \end{center}
&
\begin{center} \begin{tikzpicture}[baseline=(current bounding box.center), line width=2.2, scale=1,line cap=round]
    \draw[black] (0,0) to [bend right=15*-1] (0.8,0.4);
    \draw[black] (0.8,0.4) to [bend right=100*-1] (0.8,-0.4);
    \draw[black] (0.8,-0.4) to [bend right=15*-1] (0,0);
    \draw[black] (0,0) to [bend left=15*-1] (-0.8,0.4);
    \draw[black] (-0.8,0.4) to [bend left=100*-1] (-0.8,-0.4);
    \draw[black] (-0.8,-0.4) to [bend left=15*-1] (0,0);
    \draw[black, line width=1] (0,0) -- (-.3,.95);
    \draw[black, line width=1] (0,0) -- (.3,.95);
    \draw[black,dashed] (0,-1.1) -- (0,-.1);
	\node [black] at (0.5,.7) {$p_2$};
	\node [black] at (-0.5,.7) {$p_1$};
	\node [black] at (.34,-.9) {$p_3$};
	\node [black] at (1.35,0) {$m_t$};
	\node [black] at (-1.3,0) {$m_t$};
\end{tikzpicture} \end{center}
\\*[-.8cm]
\begin{center} $I^{(1)}_{0, 1, 2, 0, 2, 0, 0}$ \end{center} &
\begin{center} $I^{(1)}_{3, 0, 0, 0, 2, 0, 0}$ \end{center}
\end{longtable}
\end{center}

\vspace{-0.5cm}

\subsection{Family 2}

This family is spanned by thirty integrals and is defined by the propagators 
\begin{align}
D^{(2)}_i &= \{\, k_1^2 - m_t^2\, ,\, (k_1 + p_1)^2 - m_t^2\, ,\, (k_1 - p_2)^2 - m_t^2\, ,\, \nonumber \\ 
&\qquad(k_1 + k_2)^2 - M^2\, ,\, (k_2 - p_1)^2 - m_t^2\, ,\, (k_2 + p_2)^2 - m_t^2\, ,\, k_2^2 \,\} \, . \nonumber 
\end{align}
\begin{center}
\vspace{-1.1cm}
\begin{longtable}{p{8cm} | p{8cm}}
\begin{center} \begin{tikzpicture}[baseline=(current bounding box.center), line width=2.2, scale=1,line cap=round]
    \draw[black] (-2,-1) -- node[left] {$m_t$} (-2,1) -- node[above] {$m_t$} (0,1);
    \draw[black] (2.1,0) -- node[below,xshift=0.3cm,yshift=.05cm] {$m_t$} (0,-1) -- node[below] {$m_t$} (-2,-1) -- (-2,1);
    \draw[black] (0,1) to node[above,xshift=0.3cm,yshift=-.1cm]{$m_t$} (2.1,0);
    \draw[black] (0,1) -- node[left] {$M$} (0,-1);
    \draw[black, line width=1] (-2,-1) -- (-2.85,-1.85);
    \draw[black, line width=1] (-2,1) -- (-2.85,1.85);
    \draw[black,dashed] (2.1,0) -- (3.3,0);
	\node [black] at (-2.3,-1.8) {$p_1$};
	\node [black] at (-2.3,1.8) {$p_2$};
	\node [black] at (3,-.3) {$p_3$};
\end{tikzpicture} \end{center}
&
\begin{center} \begin{tikzpicture}[baseline=(current bounding box.center), line width=2.2, scale=1,line cap=round]
    \draw[black] (-2,0) -- node[above,xshift=-.1cm] {$m_t$} (0,1);
    \draw[black] (2.1,0) -- node[below,xshift=0.3cm,yshift=.05cm] {$m_t$} (0,-1);
    \draw[black] (0,-1) -- node[below,xshift=-.1cm] {$m_t$} (-2,0);
    \draw[black] (0,1) to node[above,xshift=0.3cm,yshift=-.1cm]{$m_t$} (2.1,0);
    \draw[black] (0,1) -- node[left] {$M$} (0,-1);
    \draw[black, line width=1] (-2,0) -- (-3,-.85);
    \draw[black, line width=1] (-2,0) -- (-3,.85);
    \draw[black,dashed] (2.1,0) -- (3.3,0);
	\node [black] at (-2.6,-1) {$p_1$};
	\node [black] at (-2.6,1) {$p_2$};
	\node [black] at (3,-.3) {$p_3$};
\end{tikzpicture} \end{center}
\\*[-.8cm]
\begin{center} $I^{(2)}_{1, 1, 1, 1, 1, 1, 0}$ \end{center} &
\begin{center} $I^{(2)}_{0, 1, 1, 2, 1, 1, 0}$ \end{center} \\ \hline
\begin{center} \begin{tikzpicture}[baseline=(current bounding box.center), line width=2.2, scale=1,line cap=round]
    \draw[black] (-2,0) -- node[above,xshift=-.1cm] {$m_t$} (0,1);
    \draw[black] (2.1,0) -- node[below,xshift=0.3cm,yshift=.05cm] {$m_t$} (0,-1);
    \draw[black] (0,-1) -- node[below,xshift=-.1cm] {$m_t$} (-2,0);
    \draw[black] (0,1) to node[above,xshift=0.3cm,yshift=-.1cm]{$m_t$} (2.1,0);
    \draw[black] (0,1) -- node[left] {$M$} (0,-1);
    \draw[black, line width=1] (0,1) -- (0,2);
    \draw[black, line width=1] (-2,0) -- (-3.3,0);
    \draw[black,dashed] (2.1,0) -- (3.3,0);
	\node [black] at (-2.9,-.3) {$p_1$};
	\node [black] at (-.3,1.7) {$p_2$};
	\node [black] at (3,-.3) {$p_3$};
\end{tikzpicture} \end{center}
&
\begin{center} \begin{tikzpicture}[baseline=(current bounding box.center), line width=2.2, scale=1,line cap=round]
    \draw[black] (-2,0) -- node[above,xshift=-.1cm] {$m_t$} (0,1);
    \draw[black] (2.1,0) to [bend left] node[right,xshift=.1cm,yshift=-.1cm] {$m_t$} (0,-1);
    \draw[black] (2.1,0) to [bend right] node[left,,xshift=-.1cm,yshift=.1cm] {$M$} (0,-1);
    \draw[black] (0,-1) -- node[below,xshift=-.1cm] {$m_t$} (-2,0);
    \draw[black] (0,1) to node[above,xshift=0.3cm,yshift=-.1cm]{$m_t$} (2.1,0);
    \draw[black, line width=1] (0,1) -- (0,2);
    \draw[black, line width=1] (-2,0) -- (-3.3,0);
    \draw[black,dashed] (2.1,0) -- (3.3,0);
	\node [black] at (-2.9,-.3) {$p_1$};
	\node [black] at (-.3,1.7) {$p_2$};
	\node [black] at (3,-.3) {$p_3$};
\end{tikzpicture} \end{center}
\\*[-.8cm]
\begin{center} $I^{(2)}_{1, 1, 0, 1, 1, 2, 0}$\ , \ $I^{(2)}_{1, 1, 0, 1, 2, 1, 0}$\ , \ $I^{(2)}_{1, 1, 0, 1, 1, 1, 0}$ \end{center} &
\begin{center} $I^{(2)}_{1, 1, 1, 1, 1, 0, 0}$ \end{center} \\ \hline
\begin{center} \begin{tikzpicture}[baseline=(current bounding box.center), line width=2.2, scale=1,line cap=round]
    \draw[black] (2.1,0) -- node[below,xshift=0.3cm,yshift=.05cm] {$m_t$} (0,-1);
    \draw[black] (0,1) to node[above,xshift=0.3cm,yshift=-.1cm]{$m_t$} (2.1,0);
    \draw[black] (0,1) to [bend left] node[right] {$M$} (0,-1);
    \draw[black] (0,1) to [bend right] node[left] {$m_t$} (0,-1);
    \draw[black, line width=1] (2.1,0) -- (3,-.85);
    \draw[black, line width=1] (2.1,0) -- (3,.85);
    \draw[black, dashed] (0,-1) -- (-.7,-1.85);
	\node [black] at (2.6,-1) {$p_1$};
	\node [black] at (2.6,1) {$p_2$};
	\node [black] at (-1,-1.6) {$p_3$};
\end{tikzpicture} \end{center}
&
\begin{center} \begin{tikzpicture}[baseline=(current bounding box.center), line width=2.2, scale=1,line cap=round]
    \draw[black] (2.1,0) -- node[below,xshift=0.3cm,yshift=.05cm] {$m_t$} (0,-1);
    \draw[black] (0,1) to node[above,xshift=0.3cm,yshift=-.1cm]{$m_t$} (2.1,0);
    \draw[black] (0,1) to [bend left] node[right] {$M$} (0,-1);
    \draw[black] (0,1) to [bend right] node[left] {$m_t$} (0,-1);
    \draw[black, dashed] (0,-1) -- (-.7,-1.85);
    \draw[black, line width=1] (0,1) -- (-.7,1.85);
    \draw[black, line width=1] (2.1,0) -- (3.3,0);
	\node [black] at (-1,1.6) {$p_1$};
	\node [black] at (-1,-1.6) {$p_3$};
	\node [black] at (3,-.3) {$p_2$};
\end{tikzpicture} \end{center}
\\*[-.8cm]
\begin{center} $I^{(2)}_{0, 1, 2, 1, 2, 0, 0}$ \end{center} &
\begin{center} $I^{(2)}_{1, 0, 1, 1, 2, 0, 0}$\ , \ $I^{(2)}_{1, 0, 1, 2, 1, 0, 0}$ \end{center} \\ \hline
\begin{center} \begin{tikzpicture}[baseline=(current bounding box.center), line width=2.2, scale=1,line cap=round]
    \draw[black] (0,0) to node[below,yshift=.05cm]{$m_t$} (2,0);
    \draw[black] (0,0) to [bend left=70*-1] node[below,yshift=.05cm]{$m_t$} (2,0);
    \draw[black] (0,0) to [bend right=45*-1] node[above,yshift=-.05cm]{$M$} (2,0);
    \draw[black, line width=1] (0,0) -- (-1,-.85);
    \draw[black, line width=1] (0,0) -- (-1,.85);
    \draw[black,dashed] (2.1,0) -- (3.3,0);
	\node [black] at (-0.6,-1) {$p_1$};
	\node [black] at (-0.6,1) {$p_2$};
	\node [black] at (3,-.3) {$p_3$};
\end{tikzpicture} \end{center}
&
\begin{center} \begin{tikzpicture}[baseline=(current bounding box.center), line width=2.2, scale=1,line cap=round]
    \draw[black] (-2,0) -- node[above,xshift=-.1cm] {$m_t$} (0,1);
    \draw[black, line width=1] (2.1,0) to [bend left] node[right,xshift=.1cm,yshift=-.1cm] {} (0,-1);
    \draw[black] (2.1,0) to [bend right] node[left,,xshift=-.1cm,yshift=.1cm] {$M$} (0,-1);
    \draw[black] (0,-1) -- node[below,xshift=-.1cm] {$m_t$} (-2,0);
    \draw[black] (0,1) to node[above,xshift=0.3cm,yshift=-.1cm]{$m_t$} (2.1,0);
    \draw[black, dashed] (0,1) -- (0,2);
    \draw[black, line width=1] (-2,0) -- (-3.3,0);
    \draw[black, line width=1] (2.1,0) -- (3.3,0);
	\node [black] at (-2.9,-.3) {$p_1$};
	\node [black] at (-.3,1.7) {$p_3$};
	\node [black] at (3,-.3) {$p_2$};
\end{tikzpicture} \end{center}
\\*[-.8cm]
\begin{center} $I^{(2)}_{0, 0, 1, 1, 1, 0, -1}$\ , \ $I^{(2)}_{0, 0, 2, 1, 1, 0, 0}$\ , \ $I^{(2)}_{0, 0, 1, 1, 1, 0, 0}$ \end{center} &
\begin{center} $I^{(2)}_{1, 1, 1, 1, 0, 0, 1}$ \end{center} \\ \hline
\begin{center} \begin{tikzpicture}[baseline=(current bounding box.center), line width=2.2, scale=1,line cap=round]
    \draw[black] (2,0) to [bend left] node[below] {$m_t$} (0,0);
    \draw[black] (0,0) to [bend left] node[above,yshift=-.05cm]{$m_t$} (2,0);
    \draw[black] (-2,1) to node[above, yshift=.05cm] {$m_t$} (0,0);
    \draw[black] (-2,-1) to node[below, yshift=-.1cm] {$m_t$} (0,0);
    \draw[black] (-2,-1) to node[left] {$m_t$} (-2,1);
    \draw[black,dashed] (2.1,0) -- (3.3,0);
    \draw[black, line width=1] (-2,-1) -- (-2.85,-1.85);
    \draw[black, line width=1] (-2,1) -- (-2.85,1.85);
	\node [black] at (-2.3,-1.8) {$p_1$};
	\node [black] at (-2.3,1.8) {$p_2$};
	\node [black] at (3,-.3) {$p_3$};
\end{tikzpicture} \end{center}
&
\begin{center} \begin{tikzpicture}[baseline=(current bounding box.center), line width=2.2, scale=1,line cap=round]
    \draw[black] (2.1,0) -- node[below,xshift=0.3cm,yshift=.05cm] {$m_t$} (0,-1);
    \draw[black] (0,1) to node[above,xshift=0.3cm,yshift=-.1cm]{$m_t$} (2.1,0);
    \draw[black] (0,1) to [bend left] node[right] {$M$} (0,-1);
    \draw[black] (0,1) to [bend right] node[left] {$m_t$} (0,-1);
    \draw[black, line width=1] (0,-1) -- (-.7,-1.85);
    \draw[black, line width=1] (0,1) -- (-.7,1.85);
    \draw[black,dashed] (2.1,0) -- (3.3,0);
	\node [black] at (-1,1.6) {$p_2$};
	\node [black] at (-1,-1.6) {$p_1$};
	\node [black] at (3,-.3) {$p_3$};
\end{tikzpicture} \end{center}
\\*[-.8cm]
& \begin{center} $I^{(2)}_{1, -1, 0, 1, 1, 1, 0}$\ , \ $I^{(2)}_{1, 0, 0, 1, 1, 1, -2}$ \end{center} \\[-1.6cm]
\begin{center} $I^{(2)}_{1, 1, 1, 0, 1, 1, 0}$ \end{center} & \\[-1.6cm]
& \begin{center} $I^{(2)}_{1, 0, 0, 1, 1, 1, -1}$\ , \ $I^{(2)}_{1, 0, 0, 1, 1, 1, 0}$ \end{center} \\ \hline
\begin{center} \begin{tikzpicture}[baseline=(current bounding box.center), line width=2.2, scale=1,line cap=round]
    \draw[black] (2.1,0) -- node[below,xshift=0.3cm,yshift=.05cm] {$m_t$} (0,-1);
    \draw[black] (0,1) to node[above,xshift=0.3cm,yshift=-.1cm]{$m_t$} (2.1,0);
    \draw[black] (0,1) to [bend left] node[right] {$M$} (0,-1);
    \draw[black, line width=1] (0,1) to [bend right] (0,-1);
    \draw[black, line width=1] (0,-1) -- (-.7,-1.85);
    \draw[black, line width=1] (0,1) -- (-.7,1.85);
    \draw[black,dashed] (2.1,0) -- (3.3,0);
	\node [black] at (-1,1.6) {$p_2$};
	\node [black] at (-1,-1.6) {$p_1$};
	\node [black] at (3,-.3) {$p_3$};
\end{tikzpicture} \end{center}
&
\begin{center} \begin{tikzpicture}[baseline=(current bounding box.center), line width=2.2, scale=1,line cap=round]
    \draw[black] (2,0) to [bend left] node[below] {$m_t$} (0,0);
    \draw[black] (0,0) to [bend left] node[above,yshift=-.05cm]{$m_t$} (2,0);
    \draw[black] (-2,0) to [bend left] node[above,yshift=-.05cm]{$m_t$} (0,0);
    \draw[black] (-2,0) to [bend right] node[below] {$m_t$} (0,0);
    \draw[black, line width=1] (-2,0) -- (-3,-.85);
    \draw[black, line width=1] (-2,0) -- (-3,.85);
    \draw[black,dashed] (2.1,0) -- (3.3,0);
	\node [black] at (-2.6,-1) {$p_1$};
	\node [black] at (-2.6,1) {$p_2$};
	\node [black] at (3,-.3) {$p_3$};
\end{tikzpicture} \end{center}
\\*[-.8cm]
\begin{center} $I^{(2)}_{0, 1, 1, 3, 0, 0, 1}$\ , \ $I^{(2)}_{0, 1, 2, 2, 0, 0, 1}$\ , \ $I^{(2)}_{0, 1, 1, 2, 0, 0, 1}$ \end{center} &
\begin{center} $I^{(2)}_{0, 2, 1, 0, 1, 1, 0}$ \end{center} \\ \hline
\begin{center} \begin{tikzpicture}[baseline=(current bounding box.center), line width=2.2, scale=1,line cap=round]
    \draw[black] (2.1,0) -- node[below,xshift=0.3cm,yshift=.05cm] {$m_t$} (0,-1);
    \draw[black] (0,1) to node[above,xshift=0.3cm,yshift=-.1cm]{$m_t$} (2.1,0);
    \draw[black] (0,1) to node[left] {$m_t$} (0,-1);
    \draw[black, line width=1] (0,-1) -- (-.7,-1.85);
    \draw[black, line width=1] (0,1) -- (-.7,1.85);
    \draw[black] (2.1,0) to [bend right=45*-1] (2.5,0.4);
    \draw[black] (2.5,0.4) to [bend right=45*-1] (2.9,0);
    \draw[black] (2.9,0) to [bend right=45*-1] (2.5,-0.4);
    \draw[black] (2.5,-0.4) to [bend right=45*-1] (2.1,0);
    \draw[black,dashed] (1.8,-1.1) -- (2.04,-.1);
	\node [black] at (2.2,-.9) {$p_3$};
	\node [black] at (-1,1.6) {$p_2$};
	\node [black] at (-1,-1.6) {$p_1$};
	\node [black] at (3.23,0) {$m_t$};
\end{tikzpicture} \end{center}
&
\begin{center} \begin{tikzpicture}[baseline=(current bounding box.center), line width=2.2, scale=1,line cap=round]
    \draw[black] (2.1,0) -- node[below,xshift=0.3cm,yshift=.05cm] {$m_t$} (0,-1);
    \draw[black] (0,1) to node[above,xshift=0.3cm,yshift=-.1cm]{$m_t$} (2.1,0);
    \draw[black] (0,1) to node[left] {$m_t$} (0,-1);
    \draw[black, line width=1] (0,-1) -- (-.7,-1.85);
    \draw[black, line width=1] (0,1) -- (-.7,1.85);
    \draw[black] (2.1,0) to [bend right=45*-1] (2.5,0.4);
    \draw[black] (2.5,0.4) to [bend right=45*-1] (2.9,0);
    \draw[black] (2.9,0) to [bend right=45*-1] (2.5,-0.4);
    \draw[black] (2.5,-0.4) to [bend right=45*-1] (2.1,0);
    \draw[black,dashed] (1.8,-1.1) -- (2.04,-.1);
	\node [black] at (2.2,-.9) {$p_3$};
	\node [black] at (-1,1.6) {$p_2$};
	\node [black] at (-1,-1.6) {$p_1$};
	\node [black] at (3.23,0) {$M$};
\end{tikzpicture} \end{center}
\\*[-.8cm]
\begin{center} $I^{(2)}_{1, 1, 1, 0, 2, 0, 0}$ \end{center} &
\begin{center} $I^{(2)}_{1, 1, 1, 2, 0, 0, 0}$ \end{center} \\ \hline
\begin{center} \begin{tikzpicture}[baseline=(current bounding box.center), line width=2.2, scale=1,line cap=round]
    \draw[black, line width=1] (2,0) to (0,0);
    \draw[black] (0,0) to [bend left=45*-1] node[below,yshift=-.05cm]{$m_t$} (2,0);
    \draw[black] (0,0) to [bend right=45*-1] node[above,yshift=-.05cm]{$M$} (2,0);
    \draw[black, line width=1] (2,0) -- (3.3,0);
    \draw[black, line width=1] (2,0) -- (3,.85);
    \draw[black,dashed] (3,-.85) -- (2,0);
	\node [black] at (3.1,-.3) {$p_2$};
	\node [black] at (2.6,1) {$p_1$};
	\node [black] at (2.6,-1) {$p_3$};
\end{tikzpicture} \end{center}
&
\begin{center} \begin{tikzpicture}[baseline=(current bounding box.center), line width=2.2, scale=1,line cap=round]
    \draw[black, line width=1] (2,0) to (0,0);
    \draw[black] (0,0) to node[below,yshift=.05cm]{$m_t$} (2,0);
    \draw[black] (0,0) to [bend left=70*-1] node[below,yshift=.05cm]{$m_t$} (2,0);
    \draw[black] (0,0) to [bend right=45*-1] node[above,yshift=-.05cm]{$M$} (2,0);
    \draw[black, line width=1] (0,0) -- (-1,0);
    \draw[black, line width=1] (2,0) -- (3,.85);
    \draw[black,dashed] (3,-.85) -- (2,0);
	\node [black] at (-0.6,-.3) {$p_1$};
	\node [black] at (2.6,1) {$p_2$};
	\node [black] at (2.6,-1) {$p_3$};
\end{tikzpicture} \end{center}
\\*[-.8cm]
\begin{center} $I^{(2)}_{1, 0, 0, 2, 0, 0, 1}$ \end{center} &
\begin{center} $I^{(2)}_{1, 0, 0, 1, 2, 0, 0}$ \end{center} \\ \hline
\begin{center} \begin{tikzpicture}[baseline=(current bounding box.center), line width=2.2, scale=1,line cap=round]
    \draw[black] (2,0) to [bend left] node[below] {$m_t$} (0,0);
    \draw[black] (0,0) to [bend left] node[above,yshift=-.05cm]{$m_t$} (2,0);
    \draw[black] (2,0) to [bend right=45*-1] (2.4,0.4);
    \draw[black] (2.4,0.4) to [bend right=45*-1] (2.8,0);
    \draw[black] (2.8,0) to [bend right=45*-1] (2.4,-0.4);
    \draw[black] (2.4,-0.4) to [bend right=45*-1] (2,0);
    \draw[black, line width=1] (0,0) -- (-1,-.85);
    \draw[black, line width=1] (0,0) -- (-1,.85);
    \draw[black,dashed] (1.8,-1.1) -- (1.94,-.1);
	\node [black] at (-0.6,-1) {$p_1$};
	\node [black] at (-0.6,1) {$p_2$};
	\node [black] at (2.2,-.9) {$p_3$};
	\node [black] at (3.13,0) {$m_t$};
\end{tikzpicture} \end{center}
&
\begin{center} \begin{tikzpicture}[baseline=(current bounding box.center), line width=2.2, scale=1,line cap=round]
    \draw[black] (2,0) to [bend left] node[below] {$m_t$} (0,0);
    \draw[black] (0,0) to [bend left] node[above,yshift=-.05cm]{$m_t$} (2,0);
    \draw[black] (2,0) to [bend right=45*-1] (2.4,0.4);
    \draw[black] (2.4,0.4) to [bend right=45*-1] (2.8,0);
    \draw[black] (2.8,0) to [bend right=45*-1] (2.4,-0.4);
    \draw[black] (2.4,-0.4) to [bend right=45*-1] (2,0);
    \draw[black, line width=1] (0,0) -- (-1,-.85);
    \draw[black, line width=1] (0,0) -- (-1,.85);
    \draw[black,dashed] (1.8,-1.1) -- (1.94,-.1);
	\node [black] at (-0.6,-1) {$p_1$};
	\node [black] at (-0.6,1) {$p_2$};
	\node [black] at (2.2,-.9) {$p_3$};
	\node [black] at (3.13,0) {$M$};
\end{tikzpicture} \end{center}
\\*[-.8cm]
\begin{center} $I^{(2)}_{0, 1, 2, 0, 2, 0, 0}$ \end{center} &
\begin{center} $I^{(2)}_{0, 1, 2, 2, 0, 0, 0}$ \end{center} \\ \hline
\begin{center} \begin{tikzpicture}[baseline=(current bounding box.center), line width=2.2, scale=1,line cap=round]
    \draw[black] (0,0) to [bend right=15*-1] (0.8,0.4);
    \draw[black] (0.8,0.4) to [bend right=100*-1] (0.8,-0.4);
    \draw[black] (0.8,-0.4) to [bend right=15*-1] (0,0);
    \draw[black] (0,0) to [bend left=15*-1] (-0.8,0.4);
    \draw[black] (-0.8,0.4) to [bend left=100*-1] (-0.8,-0.4);
    \draw[black] (-0.8,-0.4) to [bend left=15*-1] (0,0);
    \draw[black, line width=1] (0,0) -- (-.3,.95);
    \draw[black, line width=1] (0,0) -- (.3,.95);
    \draw[black,dashed] (0,-1.1) -- (0,-.1);
	\node [black] at (0.5,.7) {$p_2$};
	\node [black] at (-0.5,.7) {$p_1$};
	\node [black] at (.34,-.9) {$p_3$};
	\node [black] at (1.35,0) {$m_t$};
	\node [black] at (-1.3,0) {$M$};
\end{tikzpicture} \end{center}
&
\begin{center} \begin{tikzpicture}[baseline=(current bounding box.center), line width=2.2, scale=1,line cap=round]
    \draw[black] (0,0) to [bend right=15*-1] (0.8,0.4);
    \draw[black] (0.8,0.4) to [bend right=100*-1] (0.8,-0.4);
    \draw[black] (0.8,-0.4) to [bend right=15*-1] (0,0);
    \draw[black] (0,0) to [bend left=15*-1] (-0.8,0.4);
    \draw[black] (-0.8,0.4) to [bend left=100*-1] (-0.8,-0.4);
    \draw[black] (-0.8,-0.4) to [bend left=15*-1] (0,0);
    \draw[black, line width=1] (0,0) -- (-.3,.95);
    \draw[black, line width=1] (0,0) -- (.3,.95);
    \draw[black,dashed] (0,-1.1) -- (0,-.1);
	\node [black] at (0.5,.7) {$p_2$};
	\node [black] at (-0.5,.7) {$p_1$};
	\node [black] at (.34,-.9) {$p_3$};
	\node [black] at (1.35,0) {$m_t$};
	\node [black] at (-1.3,0) {$m_t$};
\end{tikzpicture} \end{center}
\\*[-.8cm]
\begin{center} $I^{(2)}_{2, 0, 0, 3, 0, 0, 0}$ \end{center} &
\begin{center} $I^{(2)}_{3, 0, 0, 0, 2, 0, 0}$ \end{center} 
\end{longtable}
\end{center}

\newpage

\subsection{Family 3}

The propagators that define this family are given by
\begin{align}
D^{(3)}_i &= \{\, k_1^2\, ,\, (k_1 + p_1)^2\, ,\, (k_1 - p_2)^2\, ,\, (k_1 + k_2)^2 - M^2\, ,\, \nonumber \\
&\qquad(k_2 - p_1)^2 - m_t^2\, ,\, (k_2 + p_2)^2 - m_t^2\, ,\, k_2^2 \,\} \, , \nonumber 
\end{align}
and it is spanned by fifteen integrals:
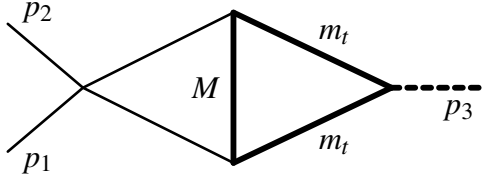
\begin{center}
\begin{longtable}{p{8cm} | p{8cm}}
\begin{center} \begin{tikzpicture}[baseline=(current bounding box.center), line width=2.2, scale=1,line cap=round]
    \draw[black,line width=1] (-2,0) -- node[above] {} (0,1);
    \draw[black] (2.1,0) -- node[below,xshift=0.3cm,yshift=.05cm] {$m_t$} (0,-1);
    \draw[black,line width=1] (0,-1) -- node[below] {} (-2,0);
    \draw[black] (0,1) to node[above,xshift=0.3cm,yshift=-.1cm]{$m_t$} (2.1,0);
    \draw[black] (0,1) -- node[left] {$M$} (0,-1);
    \draw[black, line width=1] (-2,0) -- (-3,-.85);
    \draw[black, line width=1] (-2,0) -- (-3,.85);
    \draw[black,dashed] (2.1,0) -- (3.3,0);
	\node [black] at (-2.6,-1) {$p_1$};
	\node [black] at (-2.6,1) {$p_2$};
	\node [black] at (3,-.3) {$p_3$};
\end{tikzpicture} \end{center}
&
\begin{center} \begin{tikzpicture}[baseline=(current bounding box.center), line width=2.2, scale=1,line cap=round]
    \draw[black,line width=1] (-2,0) -- node[above] {} (0,1);
    \draw[black] (2.1,0) -- node[below,xshift=0.3cm,yshift=.05cm] {$m_t$} (0,-1);
    \draw[black,line width=1] (0,-1) -- node[below] {} (-2,0);
    \draw[black] (0,1) to node[above,xshift=0.3cm,yshift=-.1cm]{$m_t$} (2.1,0);
    \draw[black] (0,1) -- node[left] {$M$} (0,-1);
    \draw[black, line width=1] (0,1) -- (0,2);
    \draw[black, line width=1] (-2,0) -- (-3.3,0);
    \draw[black,dashed] (2.1,0) -- (3.3,0);
	\node [black] at (-2.9,-.3) {$p_1$};
	\node [black] at (-.3,1.7) {$p_2$};
	\node [black] at (3,-.3) {$p_3$};
\end{tikzpicture} \end{center}
\\*[-.8cm]
\begin{center} $I^{(3)}_{0, 1, 1, 1, 1, 1, 0}$ \end{center} &
\begin{center} $I^{(3)}_{1, 1, 0, 1, 1, 1, 0}$ \end{center} \\ \hline
\begin{center} \begin{tikzpicture}[baseline=(current bounding box.center), line width=2.2, scale=1,line cap=round]
    \draw[black] (2.1,0) -- node[below,xshift=0.3cm,yshift=.05cm] {$m_t$} (0,-1);
    \draw[black] (0,1) to node[above,xshift=0.3cm,yshift=-.1cm]{$m_t$} (2.1,0);
    \draw[black] (0,1) to [bend left] node[right] {$M$} (0,-1);
    \draw[black, line width=1] (0,1) to [bend right] (0,-1);
    \draw[black, line width=1] (0,-1) -- (-.7,-1.85);
    \draw[black, line width=1] (0,-1) -- (.7,-1.85);
    \draw[black,dashed] (2.1,0) -- (3.3,0);
	\node [black] at (-.94,-1.6) {$p_2$};
	\node [black] at (1,-1.6) {$p_1$};
	\node [black] at (3,-.3) {$p_3$};
\end{tikzpicture} \end{center}
& 
\begin{center} \begin{tikzpicture}[baseline=(current bounding box.center), line width=2.2, scale=1,line cap=round]
    \draw[black] (2.1,0) -- node[below,xshift=0.3cm,yshift=.05cm] {$m_t$} (0,-1);
    \draw[black] (0,1) to node[above,xshift=0.3cm,yshift=-.1cm]{$m_t$} (2.1,0);
    \draw[black] (0,1) to [bend left] node[right] {$M$} (0,-1);
    \draw[black, line width=1] (0,1) to [bend right] (0,-1);
    \draw[black, line width=1] (0,-1) -- (-.7,-1.85);
    \draw[black, line width=1] (0,1) -- (-.7,1.85);
    \draw[black,dashed] (2.1,0) -- (3.3,0);
	\node [black] at (-1,1.6) {$p_2$};
	\node [black] at (-1,-1.6) {$p_1$};
	\node [black] at (3,-.3) {$p_3$};
\end{tikzpicture} \end{center}
\\*[-.8cm]
\begin{center} $I^{(3)}_{0, 1, 0, 2, 1, 2, 0}$ \end{center} &
\begin{center} $I^{(3)}_{1, 0, 0, 3, 1, 1, 0}$\ , \ $I^{(3)}_{1, 0, 0, 2, 1, 2, 0}$\ , \ $I^{(3)}_{1, 0, 0, 2, 1, 1, 0}$ \end{center} \\ \hline
\begin{center} \begin{tikzpicture}[baseline=(current bounding box.center), line width=2.2, scale=1,line cap=round]
    \draw[black] (2,0) to [bend left] node[below] {$m_t$} (0,0);
    \draw[black] (0,0) to [bend left] node[above,yshift=-.05cm]{$m_t$} (2,0);
    \draw[black, line width=1] (-2,0) to [bend left] (0,0);
    \draw[black, line width=1] (-2,0) to [bend right] (0,0);
    \draw[black, line width=1] (-2,0) -- (-3,-.85);
    \draw[black, line width=1] (-2,0) -- (-3,.85);
    \draw[black,dashed] (2.1,0) -- (3.3,0);
	\node [black] at (-2.6,-1) {$p_1$};
	\node [black] at (-2.6,1) {$p_2$};
	\node [black] at (3,-.3) {$p_3$};
\end{tikzpicture} \end{center}
& 
\begin{center} \begin{tikzpicture}[baseline=(current bounding box.center), line width=2.2, scale=1,line cap=round]
    \draw[black] (2,0) to [bend left] node[below] {$m_t$} (0,0);
    \draw[black] (0,0) to [bend left] node[above,yshift=-.05cm]{$m_t$} (2,0);
    \draw[black, line width=1] (0,0) -- (0.5,-1);
    \draw[black, line width=1] (0,0) -- (0.5,1);
    \draw[black] (0,0) to [bend left=45*-1] (-.4,0.4);
    \draw[black] (-.4,0.4) to [bend left=45*-1] (-.8,0);
    \draw[black] (-.8,0) to [bend left=45*-1] (-.4,-0.4);
    \draw[black] (-.4,-0.4) to [bend left=45*-1] (0,0);
    \draw[black,dashed] (2.1,0) -- (3.3,0);
	\node [black] at (-1.1,0) {$M$};
	\node [black] at (0.1,-1) {$p_1$};
	\node [black] at (0.1,1) {$p_2$};
	\node [black] at (3,-.3) {$p_3$};
\end{tikzpicture} \end{center}
\\*[-.8cm]
\begin{center} $I^{(3)}_{0, 1, 1, 0, 1, 2, 0}$ \end{center} &
\begin{center} $I^{(3)}_{0, 0, 0, 2, 2, 1, 0}$ \end{center} \\ \hline
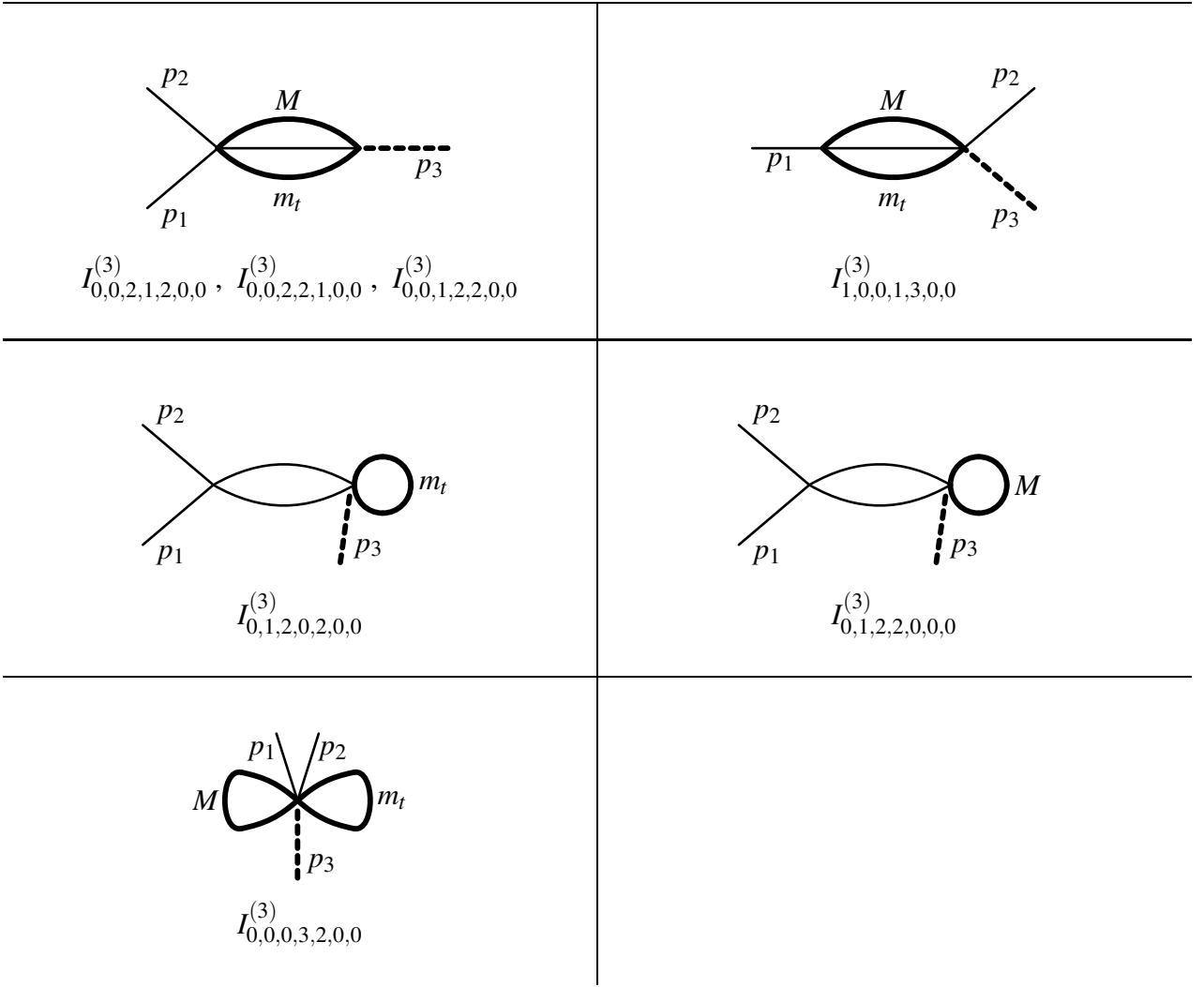
\begin{center} \begin{tikzpicture}[baseline=(current bounding box.center), line width=2.2, scale=1,line cap=round]
    \draw[black, line width=1] (2,0) to (0,0);
    \draw[black] (0,0) to [bend left=45*-1] node[below,yshift=-.05cm]{$m_t$} (2,0);
    \draw[black] (0,0) to [bend right=45*-1] node[above,yshift=-.05cm]{$M$} (2,0);
    \draw[black, line width=1] (0,0) -- (-1,-.85);
    \draw[black, line width=1] (0,0) -- (-1,.85);
    \draw[black,dashed] (2.1,0) -- (3.3,0);
	\node [black] at (-0.6,-1) {$p_1$};
	\node [black] at (-0.6,1) {$p_2$};
	\node [black] at (3,-.3) {$p_3$};
\end{tikzpicture} \end{center}
& 
\begin{center} \begin{tikzpicture}[baseline=(current bounding box.center), line width=2.2, scale=1,line cap=round]
    \draw[black, line width=1] (2,0) to (0,0);
    \draw[black] (0,0) to [bend left=45*-1] node[below,yshift=-.05cm]{$m_t$} (2,0);
    \draw[black] (0,0) to [bend right=45*-1] node[above,yshift=-.05cm]{$M$} (2,0);
    \draw[black, line width=1] (0,0) -- (-1,0);
    \draw[black, line width=1] (2,0) -- (3,.85);
    \draw[black,dashed] (3,-.85) -- (2,0);
	\node [black] at (-0.6,-.2) {$p_1$};
	\node [black] at (2.6,1) {$p_2$};
	\node [black] at (2.6,-1) {$p_3$};
\end{tikzpicture} \end{center}
\\*[-.8cm]
\begin{center} $I^{(3)}_{0, 0, 2, 1, 2, 0, 0}$\ , \ $I^{(3)}_{0, 0, 2, 2, 1, 0, 0}$\ , \ $I^{(3)}_{0, 0, 1, 2, 2, 0, 0}$ \end{center} &
\begin{center} $I^{(3)}_{1, 0, 0, 1, 3, 0, 0}$ \end{center} \\ \hline
\begin{center} \begin{tikzpicture}[baseline=(current bounding box.center), line width=2.2, scale=1,line cap=round]
    \draw[black, line width=1] (2,0) to [bend left] (0,0);
    \draw[black, line width=1] (0,0) to [bend left] (2,0);
    \draw[black] (2,0) to [bend right=45*-1] (2.4,0.4);
    \draw[black] (2.4,0.4) to [bend right=45*-1] (2.8,0);
    \draw[black] (2.8,0) to [bend right=45*-1] (2.4,-0.4);
    \draw[black] (2.4,-0.4) to [bend right=45*-1] (2,0);
    \draw[black, line width=1] (0,0) -- (-1,-.85);
    \draw[black, line width=1] (0,0) -- (-1,.85);
    \draw[black,dashed] (1.8,-1.1) -- (1.94,-.1);
	\node [black] at (-0.6,-1) {$p_1$};
	\node [black] at (-0.6,1) {$p_2$};
	\node [black] at (2.2,-.9) {$p_3$};
	\node [black] at (3.13,0) {$m_t$};
\end{tikzpicture} \end{center}
& 
\begin{center} \begin{tikzpicture}[baseline=(current bounding box.center), line width=2.2, scale=1,line cap=round]
    \draw[black, line width=1] (2,0) to [bend left] (0,0);
    \draw[black, line width=1] (0,0) to [bend left] (2,0);
    \draw[black] (2,0) to [bend right=45*-1] (2.4,0.4);
    \draw[black] (2.4,0.4) to [bend right=45*-1] (2.8,0);
    \draw[black] (2.8,0) to [bend right=45*-1] (2.4,-0.4);
    \draw[black] (2.4,-0.4) to [bend right=45*-1] (2,0);
    \draw[black, line width=1] (0,0) -- (-1,-.85);
    \draw[black, line width=1] (0,0) -- (-1,.85);
    \draw[black,dashed] (1.8,-1.1) -- (1.94,-.1);
	\node [black] at (-0.6,-1) {$p_1$};
	\node [black] at (-0.6,1) {$p_2$};
	\node [black] at (2.2,-.9) {$p_3$};
	\node [black] at (3.1,0) {$M$};
\end{tikzpicture} \end{center}
\\*[-.8cm]
\begin{center} $I^{(3)}_{0, 1, 2, 0, 2, 0, 0}$ \end{center} &
\begin{center} $I^{(3)}_{0, 1, 2, 2, 0, 0, 0}$ \end{center} \\ \hline
\begin{center} \begin{tikzpicture}[baseline=(current bounding box.center), line width=2.2, scale=1,line cap=round]
    \draw[black] (0,0) to [bend right=15*-1] (0.8,0.4);
    \draw[black] (0.8,0.4) to [bend right=100*-1] (0.8,-0.4);
    \draw[black] (0.8,-0.4) to [bend right=15*-1] (0,0);
    \draw[black] (0,0) to [bend left=15*-1] (-0.8,0.4);
    \draw[black] (-0.8,0.4) to [bend left=100*-1] (-0.8,-0.4);
    \draw[black] (-0.8,-0.4) to [bend left=15*-1] (0,0);
    \draw[black, line width=1] (0,0) -- (-.3,.95);
    \draw[black, line width=1] (0,0) -- (.3,.95);
    \draw[black,dashed] (0,-1.1) -- (0,-.1);
	\node [black] at (0.5,.7) {$p_2$};
	\node [black] at (-0.5,.7) {$p_1$};
	\node [black] at (.34,-.9) {$p_3$};
	\node [black] at (1.35,0) {$m_t$};
	\node [black] at (-1.3,0) {$M$};
\end{tikzpicture} \end{center}
\\*[-.8cm]
\begin{center} $I^{(3)}_{0, 0, 0, 3, 2, 0, 0}$ \end{center}
\end{longtable}
\end{center}

\subsection{Family 4}
\label{subsec:F4}

The propagators that define this family are given by
\begin{align}
D^{(4)}_i &= \{\, k_1^2 - m_t^2\, ,\, (k_1 + p_1)^2 - m_t^2\, ,\, (k_1 - p_2)^2 - m_t^2\, ,\, \nonumber \\ 
&\qquad(k_1 + k_2)^2\, ,\, (k_2 - p_1)^2\, ,\, (k_2 + p_2)^2 - m_t^2\, ,\, k_2^2 - m_t^2 \,\} \, , \nonumber 
\end{align}
and it is spanned by five integrals:
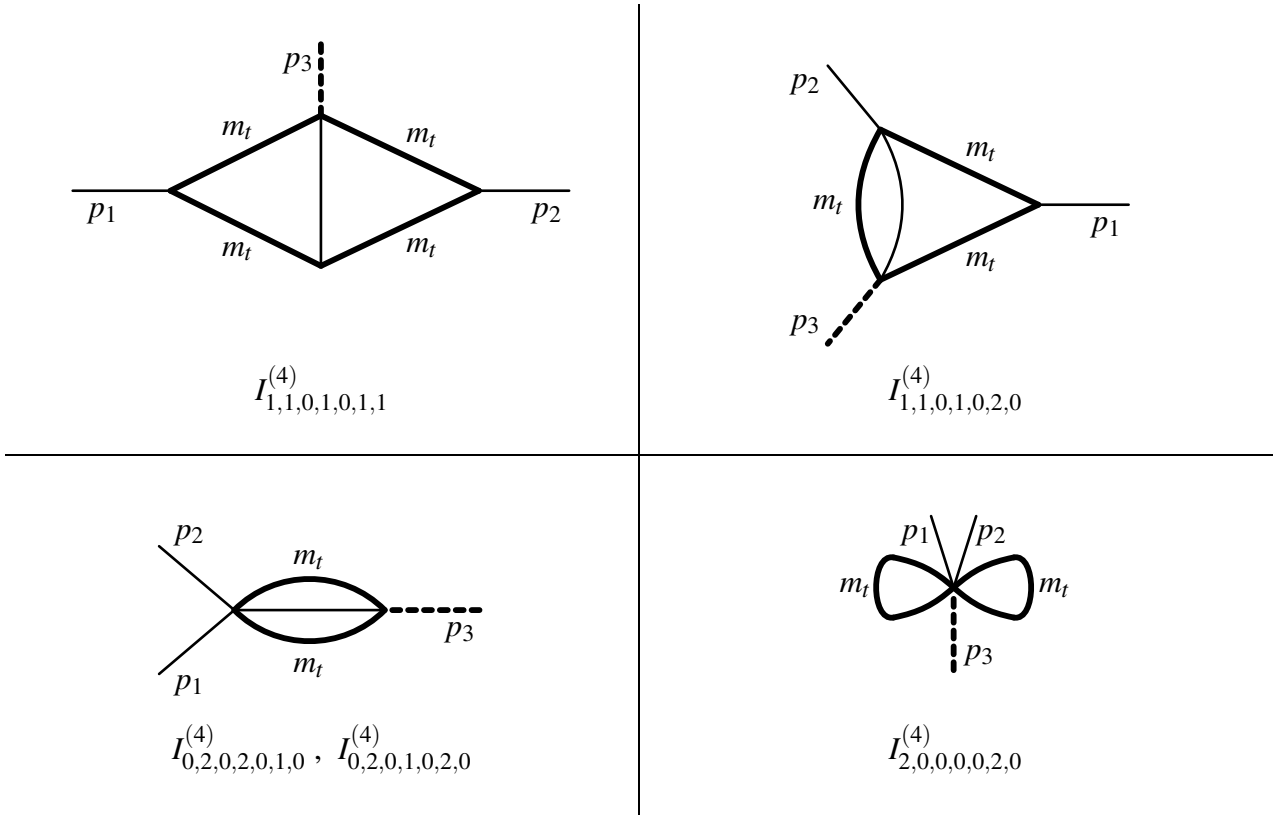
\begin{center}
\begin{longtable}{p{8cm} | p{8cm}}
\begin{center} \begin{tikzpicture}[baseline=(current bounding box.center), line width=2.2, scale=1,line cap=round]
    \draw[black] (-2,0) -- node[above,xshift=-.1cm] {$m_t$} (0,1);
    \draw[black] (2.1,0) -- node[below,xshift=0.3cm,yshift=.05cm] {$m_t$} (0,-1);
    \draw[black] (0,-1) -- node[below,xshift=-.1cm] {$m_t$} (-2,0);
    \draw[black] (0,1) to node[above,xshift=0.3cm,yshift=-.1cm]{$m_t$} (2.1,0);
    \draw[black, line width=1] (0,1) -- (0,-1);
    \draw[black, dashed] (0,1) -- (0,2);
    \draw[black, line width=1] (-2,0) -- (-3.3,0);
    \draw[black, line width=1] (2.1,0) -- (3.3,0);
	\node [black] at (-2.9,-.3) {$p_1$};
	\node [black] at (-.3,1.7) {$p_3$};
	\node [black] at (3,-.3) {$p_2$};
\end{tikzpicture} \end{center}
&
\begin{center} \begin{tikzpicture}[baseline=(current bounding box.center), line width=2.2, scale=1,line cap=round]
    \draw[black] (2.1,0) -- node[below,xshift=0.3cm,yshift=.05cm] {$m_t$} (0,-1);
    \draw[black] (0,1) to node[above,xshift=0.3cm,yshift=-.1cm]{$m_t$} (2.1,0);
    \draw[black, line width=1] (0,1) to [bend left] (0,-1);
    \draw[black] (0,1) to [bend right] node[left] {$m_t$} (0,-1);
    \draw[black, dashed] (0,-1) -- (-.7,-1.85);
    \draw[black, line width=1] (0,1) -- (-.7,1.85);
    \draw[black, line width=1] (2.1,0) -- (3.3,0);
	\node [black] at (-1,1.6) {$p_2$};
	\node [black] at (-1,-1.6) {$p_3$};
	\node [black] at (3,-.3) {$p_1$};
\end{tikzpicture} \end{center}
\\*[-.8cm]
\begin{center} $I^{(4)}_{1, 1, 0, 1, 0, 1, 1}$ \end{center} &
\begin{center} $I^{(4)}_{1, 1, 0, 1, 0, 2, 0}$ \end{center} \\ \hline
\begin{center} \begin{tikzpicture}[baseline=(current bounding box.center), line width=2.2, scale=1,line cap=round]
    \draw[black, line width=1] (2,0) to (0,0);
    \draw[black] (0,0) to [bend left=45*-1] node[below,yshift=-.05cm]{$m_t$} (2,0);
    \draw[black] (0,0) to [bend right=45*-1] node[above,yshift=-.05cm]{$m_t$} (2,0);
    \draw[black, line width=1] (0,0) -- (-1,-.85);
    \draw[black, line width=1] (0,0) -- (-1,.85);
    \draw[black,dashed] (2.1,0) -- (3.3,0);
	\node [black] at (-0.6,-1) {$p_1$};
	\node [black] at (-0.6,1) {$p_2$};
	\node [black] at (3,-.3) {$p_3$};
\end{tikzpicture} \end{center}
&
\begin{center} \begin{tikzpicture}[baseline=(current bounding box.center), line width=2.2, scale=1,line cap=round]
    \draw[black] (0,0) to [bend right=15*-1] (0.8,0.4);
    \draw[black] (0.8,0.4) to [bend right=100*-1] (0.8,-0.4);
    \draw[black] (0.8,-0.4) to [bend right=15*-1] (0,0);
    \draw[black] (0,0) to [bend left=15*-1] (-0.8,0.4);
    \draw[black] (-0.8,0.4) to [bend left=100*-1] (-0.8,-0.4);
    \draw[black] (-0.8,-0.4) to [bend left=15*-1] (0,0);
    \draw[black, line width=1] (0,0) -- (-.3,.95);
    \draw[black, line width=1] (0,0) -- (.3,.95);
    \draw[black,dashed] (0,-1.1) -- (0,-.1);
	\node [black] at (0.5,.7) {$p_2$};
	\node [black] at (-0.5,.7) {$p_1$};
	\node [black] at (.34,-.9) {$p_3$};
	\node [black] at (1.35,0) {$m_t$};
	\node [black] at (-1.3,0) {$m_t$};
\end{tikzpicture} \end{center}
\\*[-.8cm]
\begin{center} $I^{(4)}_{0, 2, 0, 2, 0, 1, 0}$\ , \ $I^{(4)}_{0, 2, 0, 1, 0, 2, 0}$ \end{center} &
\begin{center} $I^{(4)}_{2, 0, 0, 0, 0, 2, 0}$ \end{center} 
\end{longtable}
\end{center}

\subsection{Family 5}

The propagators that define this family are given by
\begin{align}
D^{(5)}_i &= \{\, k_1^2 - m_t^2\, ,\, (k_1 + p_1)^2 - m_t^2\, ,\, (k_1 - p_2)^2 - m_t^2\, ,\,  \nonumber \\ 
&\qquad(k_1 + k_2)^2 - M^2\, ,\, (k_2 - p_1)^2\, ,\, (k_2 + p_2)^2 - m_t^2\, ,\, k_2^2 - m_t^2 \,\} \, , \nonumber
\end{align}
and it is spanned by twenty-six integrals:
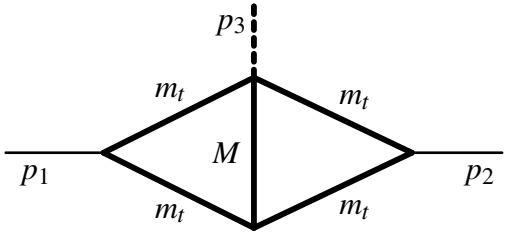
\begin{center}
\vspace{-1cm}\begin{longtable}{p{8cm} | p{8cm}}
\begin{center} \begin{tikzpicture}[baseline=(current bounding box.center), line width=2.2, scale=1,line cap=round]
    \draw[black] (-2,0) -- node[above,xshift=-.1cm] {$m_t$} (0,1);
    \draw[black] (2.1,0) -- node[below,xshift=0.3cm,yshift=.05cm] {$m_t$} (0,-1);
    \draw[black] (0,-1) -- node[below,xshift=-.1cm] {$m_t$} (-2,0);
    \draw[black] (0,1) to node[above,xshift=0.3cm,yshift=-.1cm]{$m_t$} (2.1,0);
    \draw[black] (0,1) -- node[left] {$M$} (0,-1);
    \draw[black, dashed] (0,1) -- (0,2);
    \draw[black, line width=1] (-2,0) -- (-3.3,0);
    \draw[black, line width=1] (2.1,0) -- (3.3,0);
	\node [black] at (-2.9,-.3) {$p_1$};
	\node [black] at (-.3,1.7) {$p_3$};
	\node [black] at (3,-.3) {$p_2$};
\end{tikzpicture} \end{center}
&
\begin{center} \begin{tikzpicture}[baseline=(current bounding box.center), line width=2.2, scale=1,line cap=round]
    \draw[black] (2.1,0) -- node[below,xshift=0.3cm,yshift=.05cm] {$m_t$} (0,-1);
    \draw[black] (0,1) to node[above,xshift=0.3cm,yshift=-.1cm]{$m_t$} (2.1,0);
    \draw[black] (0,1) to [bend left] node[right] {$M$} (0,-1);
    \draw[black] (0,1) to [bend right] node[left] {$m_t$} (0,-1);
    \draw[black, dashed] (0,-1) -- (-.7,-1.85);
    \draw[black, line width=1] (0,1) -- (-.7,1.85);
    \draw[black, line width=1] (2.1,0) -- (3.3,0);
	\node [black] at (-1,1.6) {$p_2$};
	\node [black] at (-1,-1.6) {$p_3$};
	\node [black] at (3,-.3) {$p_1$};
\end{tikzpicture} \end{center}
\\*[-.8cm]
\begin{center} $I^{(5)}_{1, 1, 0, 2, 0, 1, 1}$\ , \ $I^{(5)}_{1, 1, 0, 1, 0, 1, 1}$ \end{center} &
\begin{center} $I^{(5)}_{1, 1, 0, 2, 0, 1, 0}$\ , \ $I^{(5)}_{1, 1, 0, 1, 0, 2, 0}$ \end{center} \\ \hline
\begin{center} \begin{tikzpicture}[baseline=(current bounding box.center), line width=2.2, scale=1,line cap=round]
    \draw[black] (0,0) to node[below,yshift=.05cm]{$m_t$} (2,0);
    \draw[black] (0,0) to [bend left=70*-1] node[below,yshift=.05cm]{$m_t$} (2,0);
    \draw[black] (0,0) to [bend right=45*-1] node[above,yshift=-.05cm]{$M$} (2,0);
    \draw[black, line width=1] (0,0) -- (-1,-.85);
    \draw[black, line width=1] (0,0) -- (-1,.85);
    \draw[black,dashed] (2.1,0) -- (3.3,0);
	\node [black] at (-0.6,-1) {$p_1$};
	\node [black] at (-0.6,1) {$p_2$};
	\node [black] at (3,-.3) {$p_3$};
\end{tikzpicture} \end{center}
& 
\begin{center} \begin{tikzpicture}[baseline=(current bounding box.center), line width=2.2, scale=1,line cap=round]
    \draw[black] (-2,0) -- node[above,xshift=-.1cm] {$m_t$} (0,1);
    \draw[black, line width=1] (2.1,0) to [bend left] (0,-1);
    \draw[black] (2.1,0) to [bend right] node[left,,xshift=-.1cm,yshift=.1cm] {$M$} (0,-1);
    \draw[black] (0,-1) -- node[below,xshift=-.1cm] {$m_t$} (-2,0);
    \draw[black] (0,1) to node[above,xshift=0.3cm,yshift=-.1cm]{$m_t$} (2.1,0);
    \draw[black, line width=1] (0,1) -- (0,2);
    \draw[black, line width=1] (-2,0) -- (-3.3,0);
    \draw[black,dashed] (2.1,0) -- (3.3,0);
	\node [black] at (-2.9,-.3) {$p_1$};
	\node [black] at (-.3,1.7) {$p_2$};
	\node [black] at (3,-.3) {$p_3$};
\end{tikzpicture} \end{center}
\\*[-.8cm]
\begin{center} $I^{(5)}_{0, 1, 0, 1, -1, 1, 0}$\ , \ $I^{(5)}_{0, 2, 0, 1, 0, 1, 0}$\ , \ $I^{(5)}_{0, 1, 0, 1, 0, 1, 0}$ \end{center} &
\begin{center} $I^{(5)}_{1, 1, 1, 1, 1, 0, 0}$ \end{center} \\ \hline
\begin{center} \begin{tikzpicture}[baseline=(current bounding box.center), line width=2.2, scale=1,line cap=round]
    \draw[black] (-2,0) -- node[above,xshift=-.1cm] {$m_t$} (0,1);
    \draw[black] (2.1,0) to [bend left] node[right,xshift=.1cm,yshift=-.1cm] {$m_t$} (0,-1);
    \draw[black] (2.1,0) to [bend right] node[left,,xshift=-.1cm,yshift=.1cm] {$M$} (0,-1);
    \draw[black] (0,-1) -- node[below,xshift=-.1cm] {$m_t$} (-2,0);
    \draw[black] (0,1) to node[above,xshift=0.3cm,yshift=-.1cm]{$m_t$} (2.1,0);
    \draw[black, dashed] (0,1) -- (0,2);
    \draw[black, line width=1] (-2,0) -- (-3.3,0);
    \draw[black, line width=1] (2.1,0) -- (3.3,0);
	\node [black] at (-2.9,-.3) {$p_1$};
	\node [black] at (-.3,1.7) {$p_3$};
	\node [black] at (3,-.3) {$p_2$};
\end{tikzpicture} \end{center}
& 
\begin{center} \begin{tikzpicture}[baseline=(current bounding box.center), line width=2.2, scale=1,line cap=round]
    \draw[black] (2.1,0) -- node[below,xshift=0.3cm,yshift=.05cm] {$m_t$} (0,-1);
    \draw[black] (0,1) to node[above,xshift=0.3cm,yshift=-.1cm]{$m_t$} (2.1,0);
    \draw[black] (0,1) to [bend left] node[right] {$M$} (0,-1);
    \draw[black] (0,1) to [bend right] node[left] {$m_t$} (0,-1);
    \draw[black, line width=1] (0,-1) -- (-.7,-1.85);
    \draw[black, line width=1] (0,1) -- (-.7,1.85);
    \draw[black,dashed] (2.1,0) -- (3.3,0);
	\node [black] at (-1,1.6) {$p_2$};
	\node [black] at (-1,-1.6) {$p_1$};
	\node [black] at (3,-.3) {$p_3$};
\end{tikzpicture} \end{center}
\\*[-.8cm]
& \begin{center} $I^{(5)}_{0, 1, 1, 1, -1, 0, 1}$\ , \ $I^{(5)}_{-2, 1, 1, 1, 0, 0, 1}$ \end{center} \\[-1.6cm]
\begin{center} $I^{(5)}_{1, 1, 1, 1, 0, 0, 1}$ \end{center} & \\[-1.6cm]
& \begin{center} $I^{(5)}_{-1, 1, 1, 1, 0, 0, 1}$\ , \ $I^{(5)}_{0, 1, 1, 1, 0, 0, 1}$ \end{center} \\ \hline
\begin{center} \begin{tikzpicture}[baseline=(current bounding box.center), line width=2.2, scale=1,line cap=round]
    \draw[black] (2.1,0) -- node[below,xshift=0.3cm,yshift=.05cm] {$m_t$} (0,-1);
    \draw[black] (0,1) to node[above,xshift=0.3cm,yshift=-.1cm]{$m_t$} (2.1,0);
    \draw[black, line width=1] (0,1) to [bend left] node[right] {} (0,-1);
    \draw[black] (0,1) to [bend right] node[left] {$M$} (0,-1);
    \draw[black, line width=1] (2.1,0) -- (3,-.85);
    \draw[black, line width=1] (2.1,0) -- (3,.85);
    \draw[black, dashed] (0,-1) -- (-.7,-1.85);
	\node [black] at (2.6,-1) {$p_1$};
	\node [black] at (2.6,1) {$p_2$};
	\node [black] at (-1,-1.6) {$p_3$};
\end{tikzpicture} \end{center}
& 
\begin{center} \begin{tikzpicture}[baseline=(current bounding box.center), line width=2.2, scale=1,line cap=round]
    \draw[black] (2.1,0) -- node[below,xshift=0.3cm,yshift=.05cm] {$m_t$} (0,-1);
    \draw[black] (0,1) to node[above,xshift=0.3cm,yshift=-.1cm]{$m_t$} (2.1,0);
    \draw[black] (0,1) to node[left] {$m_t$} (0,-1);
    \draw[black, line width=1] (0,-1) -- (-.7,-1.85);
    \draw[black, line width=1] (0,1) -- (-.7,1.85);
    \draw[black] (2.1,0) to [bend right=45*-1] (2.5,0.4);
    \draw[black] (2.5,0.4) to [bend right=45*-1] (2.9,0);
    \draw[black] (2.9,0) to [bend right=45*-1] (2.5,-0.4);
    \draw[black] (2.5,-0.4) to [bend right=45*-1] (2.1,0);
    \draw[black,dashed] (1.8,-1.1) -- (2.04,-.1);
	\node [black] at (2.2,-.9) {$p_3$};
	\node [black] at (-1,1.6) {$p_2$};
	\node [black] at (-1,-1.6) {$p_1$};
	\node [black] at (3.23,0) {$m_t$};
\end{tikzpicture} \end{center}
\\*[-.8cm]
\begin{center} $I^{(5)}_{0, 1, 2, 2, 1, 0, 0}$ \end{center} &
\begin{center} $I^{(5)}_{1, 1, 1, 0, 0, 2, 0}$ \end{center} \\ \hline
\begin{center} \begin{tikzpicture}[baseline=(current bounding box.center), line width=2.2, scale=1,line cap=round]
    \draw[black] (2.1,0) -- node[below,xshift=0.3cm,yshift=.05cm] {$m_t$} (0,-1);
    \draw[black] (0,1) to node[above,xshift=0.3cm,yshift=-.1cm]{$m_t$} (2.1,0);
    \draw[black, line width=1] (0,1) to [bend left] node[right] {} (0,-1);
    \draw[black] (0,1) to [bend right] node[left] {$M$} (0,-1);
    \draw[black, dashed] (0,-1) -- (-.7,-1.85);
    \draw[black, line width=1] (0,1) -- (-.7,1.85);
    \draw[black, line width=1] (2.1,0) -- (3.3,0);
	\node [black] at (-1,1.6) {$p_1$};
	\node [black] at (-1,-1.6) {$p_3$};
	\node [black] at (3,-.3) {$p_2$};
\end{tikzpicture} \end{center}
& 
\begin{center} \begin{tikzpicture}[baseline=(current bounding box.center), line width=2.2, scale=1,line cap=round]
    \draw[black] (2.1,0) -- node[below,xshift=0.3cm,yshift=.05cm] {$m_t$} (0,-1);
    \draw[black] (0,1) to node[above,xshift=0.3cm,yshift=-.1cm]{$m_t$} (2.1,0);
    \draw[black] (0,1) to node[left] {$m_t$} (0,-1);
    \draw[black, line width=1] (0,-1) -- (-.7,-1.85);
    \draw[black, line width=1] (0,1) -- (-.7,1.85);
    \draw[black] (2.1,0) to [bend right=45*-1] (2.5,0.4);
    \draw[black] (2.5,0.4) to [bend right=45*-1] (2.9,0);
    \draw[black] (2.9,0) to [bend right=45*-1] (2.5,-0.4);
    \draw[black] (2.5,-0.4) to [bend right=45*-1] (2.1,0);
    \draw[black,dashed] (1.8,-1.1) -- (2.04,-.1);
	\node [black] at (2.2,-.9) {$p_3$};
	\node [black] at (-1,1.6) {$p_2$};
	\node [black] at (-1,-1.6) {$p_1$};
	\node [black] at (3.23,0) {$M$};
\end{tikzpicture} \end{center}
\\*[-.8cm]
\begin{center} $I^{(5)}_{1, 0, 1, 2, 1, 0, 0}$ \end{center} &
\begin{center} $I^{(5)}_{1, 1, 1, 2, 0, 0, 0}$ \end{center} \\ \hline
\begin{center} \begin{tikzpicture}[baseline=(current bounding box.center), line width=2.2, scale=1,line cap=round]
    \draw[black, line width=1] (2,0) to (0,0);
    \draw[black] (0,0) to [bend left=45*-1] node[below,yshift=-.05cm]{$m_t$} (2,0);
    \draw[black] (0,0) to [bend right=45*-1] node[above,yshift=-.05cm]{$M$} (2,0);
    \draw[black, line width=1] (0,0) -- (-1,-.85);
    \draw[black, line width=1] (0,0) -- (-1,.85);
    \draw[black,dashed] (2.1,0) -- (3.3,0);
	\node [black] at (-0.6,-1) {$p_1$};
	\node [black] at (-0.6,1) {$p_2$};
	\node [black] at (3,-.3) {$p_3$};
\end{tikzpicture} \end{center} &
\begin{center} \begin{tikzpicture}[baseline=(current bounding box.center), line width=2.2, scale=1,line cap=round]
    \draw[black, line width=1] (2,0) to (0,0);
    \draw[black] (0,0) to node[below,yshift=.05cm]{$m_t$} (2,0);
    \draw[black] (0,0) to [bend left=70*-1] node[below,yshift=.05cm]{$m_t$} (2,0);
    \draw[black] (0,0) to [bend right=45*-1] node[above,yshift=-.05cm]{$M$} (2,0);
    \draw[black, line width=1] (0,0) -- (-1,0);
    \draw[black, line width=1] (2,0) -- (3,.85);
    \draw[black,dashed] (3,-.85) -- (2,0);
	\node [black] at (-0.6,-.3) {$p_2$};
	\node [black] at (2.6,1) {$p_1$};
	\node [black] at (2.6,-1) {$p_3$};
\end{tikzpicture} \end{center}
\\*[-.8cm]
\begin{center} $I^{(5)}_{0, 0, 2, 1, 2, 0, 0}$\ , \ $I^{(5)}_{0, 0, 1, 2, 2, 0, 0}$\ , \ $I^{(5)}_{0, 0, 2, 2, 1, 0, 0}$ \end{center} & 
\begin{center} $I^{(5)}_{1, 0, 0, 1, 0, 2, 0}$ \end{center} \\ \hline
\begin{center} \begin{tikzpicture}[baseline=(current bounding box.center), line width=2.2, scale=1,line cap=round]
    \draw[black, line width=1] (2,0) to (0,0);
    \draw[black] (0,0) to [bend left=45*-1] node[below,yshift=-.05cm]{$m_t$} (2,0);
    \draw[black] (0,0) to [bend right=45*-1] node[above,yshift=-.05cm]{$M$} (2,0);
    \draw[black, line width=1] (0,0) -- (-1,-.85);
    \draw[black, dashed] (0,0) -- (-1,.85);
    \draw[black, line width=1] (2,0) -- (3.2,0);
	\node [black] at (-0.6,-1) {$p_2$};
	\node [black] at (-0.6,1) {$p_3$};
	\node [black] at (3,-.3) {$p_1$};
\end{tikzpicture} \end{center} & 
\begin{center} \begin{tikzpicture}[baseline=(current bounding box.center), line width=2.2, scale=1,line cap=round]
    \draw[black] (2,0) to [bend left] node[below] {$m_t$} (0,0);
    \draw[black] (0,0) to [bend left] node[above,yshift=-.05cm]{$m_t$} (2,0);
    \draw[black] (2,0) to [bend right=45*-1] (2.4,0.4);
    \draw[black] (2.4,0.4) to [bend right=45*-1] (2.8,0);
    \draw[black] (2.8,0) to [bend right=45*-1] (2.4,-0.4);
    \draw[black] (2.4,-0.4) to [bend right=45*-1] (2,0);
    \draw[black, line width=1] (0,0) -- (-1,-.85);
    \draw[black, line width=1] (0,0) -- (-1,.85);
    \draw[black,dashed] (1.8,-1.1) -- (1.94,-.1);
	\node [black] at (-0.6,-1) {$p_1$};
	\node [black] at (-0.6,1) {$p_2$};
	\node [black] at (2.2,-.9) {$p_3$};
	\node [black] at (3.13,0) {$m_t$};
\end{tikzpicture} \end{center}
\\*[-.8cm]
\begin{center} $I^{(5)}_{1, 0, 0, 2, 1, 0, 0}$ \end{center} & 
\begin{center} $I^{(5)}_{0, 1, 2, 0, 0, 2, 0}$ \end{center} \\ \hline
\begin{center} \begin{tikzpicture}[baseline=(current bounding box.center), line width=2.2, scale=1,line cap=round]
    \draw[black] (2,0) to [bend left] node[below] {$m_t$} (0,0);
    \draw[black] (0,0) to [bend left] node[above,yshift=-.05cm]{$m_t$} (2,0);
    \draw[black] (2,0) to [bend right=45*-1] (2.4,0.4);
    \draw[black] (2.4,0.4) to [bend right=45*-1] (2.8,0);
    \draw[black] (2.8,0) to [bend right=45*-1] (2.4,-0.4);
    \draw[black] (2.4,-0.4) to [bend right=45*-1] (2,0);
    \draw[black, line width=1] (0,0) -- (-1,-.85);
    \draw[black, line width=1] (0,0) -- (-1,.85);
    \draw[black,dashed] (1.8,-1.1) -- (1.94,-.1);
	\node [black] at (-0.6,-1) {$p_1$};
	\node [black] at (-0.6,1) {$p_2$};
	\node [black] at (2.2,-.9) {$p_3$};
	\node [black] at (3.13,0) {$M$};
\end{tikzpicture} \end{center} &
\begin{center} \begin{tikzpicture}[baseline=(current bounding box.center), line width=2.2, scale=1,line cap=round]
    \draw[black] (0,0) to [bend right=15*-1] (0.8,0.4);
    \draw[black] (0.8,0.4) to [bend right=100*-1] (0.8,-0.4);
    \draw[black] (0.8,-0.4) to [bend right=15*-1] (0,0);
    \draw[black] (0,0) to [bend left=15*-1] (-0.8,0.4);
    \draw[black] (-0.8,0.4) to [bend left=100*-1] (-0.8,-0.4);
    \draw[black] (-0.8,-0.4) to [bend left=15*-1] (0,0);
    \draw[black, line width=1] (0,0) -- (-.3,.95);
    \draw[black, line width=1] (0,0) -- (.3,.95);
    \draw[black,dashed] (0,-1.1) -- (0,-.1);
	\node [black] at (0.5,.7) {$p_2$};
	\node [black] at (-0.5,.7) {$p_1$};
	\node [black] at (.34,-.9) {$p_3$};
	\node [black] at (1.35,0) {$m_t$};
	\node [black] at (-1.3,0) {$M$};
\end{tikzpicture} \end{center}
\\*[-.8cm]
\begin{center} $I^{(5)}_{0, 1, 2, 2, 0, 0, 0}$ \end{center} & 
\begin{center} $I^{(5)}_{2, 0, 0, 3, 0, 0, 0}$ \end{center} \\ \hline
\begin{center} \begin{tikzpicture}[baseline=(current bounding box.center), line width=2.2, scale=1,line cap=round]
    \draw[black] (0,0) to [bend right=15*-1] (0.8,0.4);
    \draw[black] (0.8,0.4) to [bend right=100*-1] (0.8,-0.4);
    \draw[black] (0.8,-0.4) to [bend right=15*-1] (0,0);
    \draw[black] (0,0) to [bend left=15*-1] (-0.8,0.4);
    \draw[black] (-0.8,0.4) to [bend left=100*-1] (-0.8,-0.4);
    \draw[black] (-0.8,-0.4) to [bend left=15*-1] (0,0);
    \draw[black, line width=1] (0,0) -- (-.3,.95);
    \draw[black, line width=1] (0,0) -- (.3,.95);
    \draw[black,dashed] (0,-1.1) -- (0,-.1);
	\node [black] at (0.5,.7) {$p_2$};
	\node [black] at (-0.5,.7) {$p_1$};
	\node [black] at (.34,-.9) {$p_3$};
	\node [black] at (1.35,0) {$m_t$};
	\node [black] at (-1.3,0) {$m_t$};
\end{tikzpicture} \end{center}
\\*[-.8cm]
\begin{center} $I^{(5)}_{3, 0, 0, 0, 0, 2, 0}$ \end{center} 
\end{longtable}
\end{center}

{\footnotesize
\bibliography{ew_ggH}
\bibliographystyle{JHEP.bst}
}

\end{document}